\documentclass[twocolumn,showpacs,preprintnumbers,amsmath,amssymb,psamsfonts]{revtex4}
\usepackage{color}
\usepackage{citesort}
\newtheorem{corollary}{Corollary}
\newtheorem{lemma}{Lemma}
\newtheorem{proposition}{Proposition}
\newtheorem{remark}{Remark}

\newcommand{\HRule}{\rule{\linewidth}{0.5pt}}
\renewcommand{\Re}{\operatorname{Re}}
\renewcommand{\Im}{\operatorname{Im}}
\renewcommand{\vec}{\operatorname{vec}}

\newcommand{\R}[1]{\ensuremath{\mathbb R}^{\,#1}{}}
\newcommand{\C}[1]{\ensuremath{\mathbb C}^{\,#1}{}}
\newcommand{\unity}{\ensuremath{{\rm 1 \negthickspace l}{}}}

\newcommand{\RED}[1]{\textcolor{red}{#1}}

\newcommand{\braket}[2]{\ensuremath{\langle #1 | #2 \rangle}{}}

\newcommand{\ad}[1]{\operatorname{ad_{#1}}}
\newcommand{\Ad}[1]{\operatorname{Ad_{#1}}}
\newcommand{\adr}{\operatorname{ad}}
\newcommand{\Adr}{\operatorname{Ad}}
\newcommand{\Adj}{\operatorname{Adj}{}}
\newcommand{\Mat}{\operatorname{Mat}{}}
\newcommand{\diag}{\operatorname{diag}{}}

\renewcommand{\mod}{\operatorname{mod}{\;}}

\newcommand{\TE}[2]{\ensuremath{T_{#1,#2}{}}}

\newcommand{\JZ}[1]{\ensuremath{J_{\nu_{#1}}^{(#1)}{}}}

\newcommand{\WigD}[3]{\ensuremath{D_{#1,\,#2}^{(#3)}}}
\newcommand{\WigDabc}[1]{\ensuremath{D^{(#1)}(\alpha,\beta,\gamma)}}

\newcommand{\WigDOOp}[1]{\ensuremath{D^{(#1)}(0,0,\phi)}}
\newcommand{\WigDabcm}[3]{\ensuremath{D_{#1,\,#2}^{(#3)}(\alpha,\beta,\gamma)}}
\newcommand{\Wigd}[3]{\ensuremath{d_{#1,\,#2}^{(#3)}}}

\newcommand{\Partial}[2]{\ensuremath{\frac{\partial{#1}}{\partial{#2}}}\,\,{}}

\newcommand{\iso}{\ensuremath{\overset{\rm iso}{=}}{}}

\newcommand{\norm}[1]{\ensuremath{\Vert #1 \Vert{}}}
\newcommand{\fnorm}[1]{\ensuremath{\Vert #1 \Vert{}}_2^{\phantom{2}}}
\newcommand{\fnormsq}[1]{\ensuremath{\Vert #1 \Vert{}}_2^2}

\newcommand{\grad}{\operatorname{grad}}
\newcommand{\tr}{\operatorname{tr}}
\newcommand{\comm}[2]{\ensuremath{[#1,#2]}}

\newcommand{\op}{
	\multiputlist(0,2)(9,0)[0,0]{{\hspace{2.5mm}\circle*{5}},{\line(1,0){15}},{\hspace{2.5mm}\circle{5}}}
	}
\newcommand{\pp}{
	\multiputlist(0,2)(9,0)[0,0]{{\hspace{2.5mm}\circle*{5}},{\line(1,0){15}},{\hspace{2.5mm}\circle*{5}}}
	}

\newcommand{\fvo}{1}
\newcommand{\fvn}[1]{+#1}
\newcommand{\fvi}[1]{-#1}
\newcommand{\fvp}[1]{\pm#1}

\newcommand{\fwn}[1]{\star#1}
\newcommand{\fwi}[1]{\star#1}
\newcommand{\Rfn}{\RED{\fvn{}}}
\newcommand{\Rfi}{\RED{\fvi{}}}

\newcommand{\cdf}{\cdot & \cdot & \cdot & \cdot}
\newcommand{\cdt}{\cdot & \cdot & \cdot}
\newcommand{\cdw}{\cdot & \cdot}
\newcommand{\cdo}{\cdot}
 
\usepackage{epic}
\usepackage{german}
\usepackage{type1cm}

\usepackage{graphicx}
\usepackage{dcolumn}
\usepackage{bm}

\begin{document}


\title{Which Quantum Evolutions Can Be Reversed by Local Unitary Operations?\\[1mm]
 Algebraic Classification and Gradient-Flow-Based Numerical Checks}

\author{T.~Schulte-Herbr{\"u}ggen}\email{tosh@ch.tum.de}
\author{A.~Sp{\"o}rl}
\affiliation{Department of Chemistry, Technical University Munich, Lichtenbergstrasse 4, D-85747 Garching, Germany}

\date{\today}


\begin{abstract}
Generalising in the sense of Hahn's spin echo,
we completely characterise those unitary propagators of effective multi-qubit interactions
that can be inverted solely by {\em local} unitary operations on $n$ qubits (spins-$\tfrac{1}{2}$). 
The subset of $U\in \mathbf{SU}(2^n)$ satisfying $U^{-1}=K_1 U K_2$
with pairs of local unitaries $K_1, K_2\in\mathbf{SU}(2)^{\otimes n}$ comprises two classes: 
in type-I, $K_1$ and $K_2$ are inverse to one another, while in type-II they are not.
{Type-I} consists of one-parameter groups 
that can jointly be inverted for all times~$t\in\R{}$ because their Hamiltonian generators satisfy 
$K H K^{-1} = \Ad K (H) = -H$. 
As all the Hamiltonians generating locally invertible unitaries of type-I
are spanned by the eigenspace associated to the eigenvalue $-1$ of the {\em local} conjugation map $\Ad K$,
this eigenspace can be given in closed algebraic form.
The relation to the root space decomposition of $\mathfrak{sl}(N,\C{})$ is pointed out.
Special cases of type-I invertible Hamiltonians are of $p$-quantum order and are analysed 
by the transformation properties of spherical tensors of order $p$.
Effective multi-qubit interaction Hamiltonians are characterised via the graphs of their coupling topology.
{Type-II} consists of pointwise locally invertible propagators, part of which can be
classified according to the symmetries of their matrix representations.
Moreover, we show gradient flows for numerically solving the decision problem whether 
a propagator is type-I or type-II invertible or not by driving the least-squares distance
$\norm{K_1 e^{-itH} K_2 - e^{+itH}}^2_2$ to zero.

\end{abstract}

\maketitle

\section*{Introduction}

Richard Feynman's seminal conjecture that quantum systems may be used to efficiently compute
and predict the behaviour of other quantum systems \cite{Fey82} has inaugurated
branches of research {\em inter alia} dedicated to Hamiltonian 
Simulation \cite{Lloyd96,AL97,Zal98,BCL+02,MVL02,JC03}.
Actually, backed by the considerations by Manin \cite{Manin99},
Bennett \cite{Bennett82} and others, it initiated efforts to explore the power
of quantum computing. Soon thereafter, sets of {\em universal} one- and two-qubit
gates were found \cite{Deu85} which allow for decomposing any unitary representation
of a quantum computational gate into elementary universal ones. 

Exploring computational complexity as well as devising timeoptimal realisations
of given quantum algorithms by admissible controls has therefore become an
issue of considerable practical interest, see e.g. \cite{Khaneja01b,PRA05}. 
In particular the number of computational
steps required to implement a quantum gate or to simulate its Hamiltonian is
a measure of the actual cost to put the gate or the simulation into practice.
A specific question is, whether the sign-inverted Hamiltonian $-H$ can be simulated with only
$+H$ and a set of given control Hamiltonians being at hand.
The work of Beth {\em et al.} has addressed this problem for pair interactions
to give bounds on the time-overhead \cite{WJB02,WRJB02,JWB02} required for doing so.
In view of effective multi-qubit interactions, here we go beyond pair interactions
and classify those Hamiltonians that allow for simulating 
$-H$ by $H$ and local controls with exact time-overhead $1$.

This is of practical relevance, because when simulating quantum systems one often faces two-part generic tasks:
(i) let certain interactions evolve while (ii) other effective multi-qubit interactions  
shall be suppressed. The latter may be achieved by decoupling, but often
it suffices that unwanted interactions cancel at a certain time, {\em e.g.} right at the end of
an experiment, which is to say they are to be refocussed by inverting them at suitable points
in time. This is important for instance to avoid undesired or dissipative coupling of
a quantum system to its environment or bath \cite{VKLabc}. Many techniques have been developed in
magnetic resonance \cite{EBW87,Shaka96} on the basis of average-Hamiltonian theory \cite{Waugh96}. 
Moreover, local inversions arise in the
context of LOCC, i.e. local operations and classical communication \cite{VC02}.

Let the operator $T(t)$ denote {\em time translation} by $t$ while $\Theta$
represents {\em time reversal}. 
Following Wigner~\cite{Wigner32,Wigner59} in these very general terms, one immediately finds
\begin{equation}\label{eqn:time-inversion}
T(t) \,\circ\, \Theta \,\circ\, T(t) \,\circ\, \Theta = \unity \Leftrightarrow %
		\Theta\,\circ\, T(t)\,\circ\, \Theta = T^{-1}(t) \equiv T(-t)\,.
\end{equation}
Now imagine time translation is accompanied by the Hamiltonian unitary evolution of some quantum 
interaction $H$ according to $U(t) = e^{-it H}$ for all $t$. Then 
Eqn.~\ref{eqn:time-inversion} turns into
\begin{equation}\label{eqn:evol-reversal}
	\Theta\,\circ\, U(t) \,\circ\,\Theta \;= \;U(-t)\quad.
\end{equation}
Clearly time reversal itself is an unphysical operation, however, there are manipulations
that bring about effective time reversal for evolutions of certain quantum interactions,
the most prominent early example of which being Hahn's spin echo~\cite{Hahn50}.
\begin{figure*}[Ht!]
\includegraphics[width=0.95\textwidth]{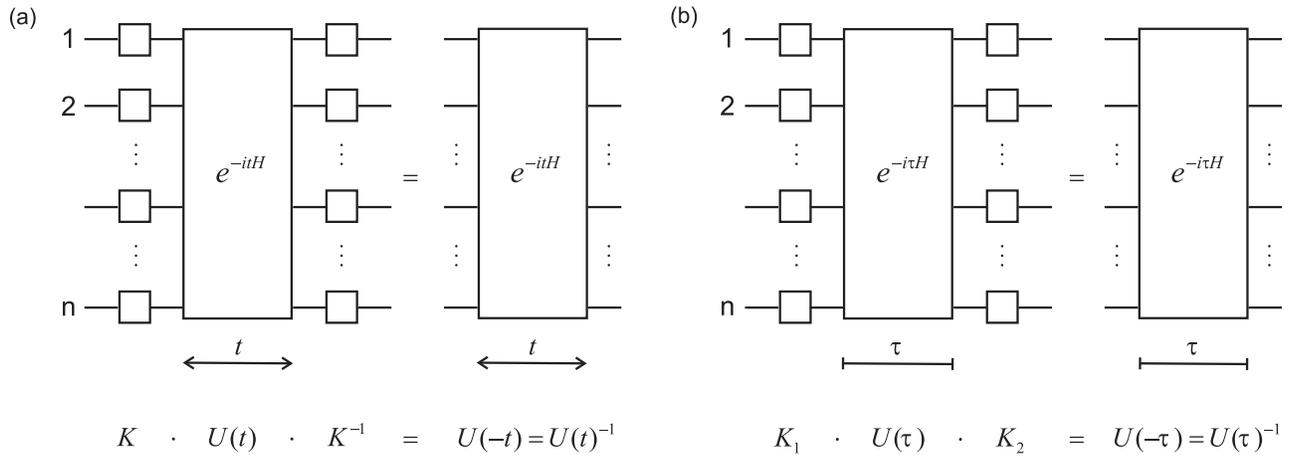}
\caption{\label{fig:inv-scheme} There are two non-trivial instances of 
locally invertible quantum evolutions: (a)
in Type-I the interaction can be refocussed by local unitary conjugation with $K\in\mathbf{SU}(2^n)$
for all times $t\in\R{}$. 
(b) Type-II locally invertible interactions can only be refocussed at specific times $\tau\in\R{}$
since $K_1$ and $K_2$ are not inverse to one another.
}
\end{figure*}

Generalising the sense of Hahn's spin echo, here we ask for which (non-trivial) Hamiltonian evolutions
effective time reversal can be obtained by {\em local} unitary operations,
in other words which Hamiltonian evolutions can be entirely refocussed by 
framing them solely with local unitaries.
As illustrated in Fig.~\ref{fig:inv-scheme} for $n$ qubits, by this we mean:
a unitary quantum propagator or gate $U := e^{-it H}\in\mathbf{SU}(2^n)$ (with non-zero $t\in\R{}$)
is invertible exclusively by {\em local unitary operations}, if
\begin{equation}
\exists K_1, K_2 \in \mathbf{SU}(2)^{\otimes n} : K_1 e^{-it H} K_2 = e^{+it H}\;.
\end{equation}
Whether or not $K_1$ and $K_2$ are inverse to one another has implications for universality, as will be
shown next.

\subsection*{Case Distinction}
\noindent
As illustrated in Fig.~\ref{fig:inv-scheme}, 
locally invertible propagators exist in two types:
\setcounter{lemma}{-1}
\begin{lemma}
Either $e^{-itH}$ is trivial and self-inverse, or (i) it is type-I invertible in the sense 
$\exists K\in \mathbf{SU}(2)^{\otimes n}: KHK^{-1}=-H$ so $K e^{-itH} K^{-1} = e^{+itH}$ jointly for all $t\in\R{}$, 
or (ii) it is type-II invertible such that at some (but not all) points $\tau$ in time 
$K_1 e^{-i\tau H} K_2 = e^{+i\tau H}$ with $K_1, K_2 \in \mathbf{SU}(2)^{\otimes n}$
and $K_2^{\phantom{1}}\neq K_1^{-1}$.
\end{lemma}
\begin{table}[Hb!]
\begin{center}
\caption{Types of locally invertible interaction evolutions}\label{tab:classes}
\begin{tabular}{l|cccc}
\hline\hline\\[-1mm]
local invertibility &\phantom{uu}& $K_2^{\phantom{1}}=K_1^{-1}$ &\phantom{uuuu}& $K_2^{\phantom{1}}\neq K_1^{-1}$\\[2mm]
\hline\\[-1mm]
jointly for all $t \in \R{}$ &&Type-I && $\{\}$ \\[2mm]
pointwise for some $\tau \in \R{}$ && self-inverse &&  Type-II \\[2mm]
\hline\hline
\end{tabular}\hspace{15mm}
\end{center}
\end{table}
Since self-inverse cases with $U^2 = \unity$ (as in the quantum computational {\sc cnot}, 
{\sc swap} and {\sc Toffoli} gates) are trivial from the point of view of inversion
they will be not discussed any further. 
In order not to bother the reader with a stumbling block, the case distinction of
Lemma~0 will be proven in Appendix~A.

\subsection*{Organisation and Notation}
Rather, for the sake of illustration, we first address the problem of type-I 
invertibility from a geometric point of view leading to Lie-algebraic terms.
A superoperator representation of the adjoint mapping then
turns out to reduce the problem 
to a simple eigenoperator calculation
that can be solved algebraically in closed form. This paves the way to discuss pair interaction
Hamiltonians. Connecting several pairs then leads to assess local invertibilty in terms of
coupling graphs in the case of multi-qubit systems. The transformation properties of
multi-qubit interactions seen as spherical tensors of different quantum orders $p$
allow for treating the problem in its normal form, namely invertibility by joint or
individual local $z$-rotations. Moreover, the quantum order $p$ relates to the roots
of the standard root-space decomposition, which gives necessary and sufficient conditions
for local invertibility. After relating the findings to Cartan-like decompositons
induced by the concurrence Cartan involution linked to time-reversal symmetry, we
finally establish the relation to global minima 
of the least-squares distance
over the restricted group of local unitaries. So invoking the norm property 
associated with the Frobenius distance,
$e^{-itH}$ is locally invertible if and only if
\begin{equation}
\underset{K_1, K_2\in SU(2)^{\otimes n}}{\min} \norm{K_1 e^{-itH} K_2 - e^{+itH}}^2_2 = 0 \quad ,
\end{equation}
which can readily be decided on numerically by a gradient flow restricited to the local
unitary group.

Type-II invertibility is first treated by establishing symmetry properties for the
matrix representations of the interaction Hamiltonians. Then a system of two coupled gradient flows
is devised to solve the problem numerically. It can be seen as a flow for the singular-value
decomposition (SVD), yet restricted to local unitaries $U$ and $V$.

Throughout the paper, we use the following notation:
${\mathbf G} := \mathbf{SU}(2^n)$, ${\mathbf K} :=  \mathbf{SU}(2)\otimes \mathbf{SU}(2) \otimes\cdots%
	\otimes\mathbf{SU}(2)=:\mathbf{SU}(2)^{\otimes n}$ for
the Lie groups of the special unitaries and local unitaries as well as
${\mathfrak g}$ and ${\mathfrak k}$ for their respective Lie algebras.
Elements of ${\mathbf G}$, ${\mathbf K}$, ${\mathfrak g}$ and ${\mathfrak k}$
are written as $G,K; g,k$ with the only obvious exception of expressing Hamiltonians by $i H\in\mathfrak g$.

\section{Locally Invertible One-Parameter Unitary Groups}
\subsection*{Prelude: Geometry}

In the single-qubit case, a spin-$\tfrac{1}{2}$ rotation $U({\bf n},\phi)$
by some angle $\phi$ about the axis
${\bf n}$ is---most intuitively---inverted by a $\pi$-rotation about
some axis ${\bf n^\bot}$ orthogonal to ${\bf n}$ according to
\begin{equation}
U({\bf n},\phi)^{-1} = U({\bf n^\bot},\pi)\; U({\bf n},\phi)\; U({\bf n^\bot},\pi)^{-1}
	\quad .
\end{equation}
Let the Lie algebra $\mathfrak{su}(2^n)$ be spanned by some orthonormal basis set $\{a_j\}$.
Then the generalisation of rotations to higher dimensions, 
e.g. $n$ qubits (spins-$\tfrac{1}{2}$)
is straightforward: replace the rotation axis by the subspace of
the Lie algebra $\mathfrak{su}(2^n)$ that is invariant under the
action of $H$ 
\begin{equation}
I_H := {\rm span}_\R{}\{ a_j \in \mathfrak{su}(2^n) \big|\; \comm{a_j}{H\,} = 0 \;\}\quad ,
\end{equation}
and consider its orthocomplement in $\mathfrak{su}(2^n)$ 
\begin{equation}
I^\bot_H := {\rm span}_\R{}\{ a_j \in \mathfrak{su}(2^n) \big|\; a_j \not\in I_H\;\}\quad .
\end{equation}
Making use of the Hilbert space structure \cite{Achieser},
every $H$ induces a specific decomposition~\footnote{for distinction from a Cartan-like decomposition, see section Algebra II} 
\begin{equation}
\mathfrak{su}(2^n) = I_H^{\phantom{\bot}}\,\oplus\,I^\bot_H \quad .
\end{equation}

\noindent
This setting already implies a particularly simple and illustrative first characterisation
of locally invertible unitaries, which, however, is not yet complete:
\begin{lemma}
For a propagator $U := e^{-it H}$ to be locally invertible for all $t$,
the orthocomplement $I^\bot_H$ to its invariant subspace in $\mathfrak{su}(2^n)$
necessarily has to comprise at least one local effective Hamiltonian $k$ with
$K := e^{-i k} \in {\bf SU}(2)^{\otimes n}$.
\end{lemma}

\noindent
{\bf Proof:}
Assume there were no local Hamiltonian $k\in I^\bot_{\rm H}$: then
all the  local unitaries $\mathbf K = {\bf SU}(2)^{\otimes n}$ would be
an invariant subgroup to ${\bf SU}(2^n)$ under the action of the one-parameter unitary group 
\mbox{$\{U = e^{-it H}\;|\; t\in \R{}\,\}$}.
In turn, there would be no local unitary to invert a propagator $U = e^{-it H}$ for all $t$.
\hfill$\blacksquare$

\begin{lemma}
For a $U = e^{-it H}$ to be locally invertible at all $t$, it is sufficient that
there is a local Hamiltonian $k$ 
in the orthocomplement $I^\bot_{\rm H}$ 
so that the double commutator of $H$ with $k$ reproduces $H$, i.e.
$[k, [k, H]]= H$.
\end{lemma}

\noindent
{\bf Proof:}
In the first place, note that ${\rm span}_\R{}\{H,k,i[H,k]\;\}
\iso \mathfrak{su}(2)$ implies $[k, [k,H]]=H$,
whereas the converse does not necessarily hold. However, the condition
$[k, [k,H]]=H$ suffices to define an analytic function
\begin{equation}
f(\phi):=e^{-i\phi k} H e^{i\phi k}
\end{equation}
with the derivatives
\begin{eqnarray}
\frac{df}{d\phi} &=& -\; e^{-i\phi k} i[k,H] e^{i\phi k}\\
\frac{d^2f}{d\phi^2} &=& -\; e^{-i\phi k} [k,[k,H]] e^{i\phi k}
 = - f(\phi)\quad .
\end{eqnarray}
The boundary conditions 
$f(0)=H$ and $\tfrac{df}{d\phi}\big|_{\phi = 0} = -i[k,H]$
allow for expressing the function $f(\phi)$ as
\begin{equation}
e^{-i\phi k} H e^{i\phi k} = H\cos\phi -i[k,H]\sin\phi
\quad .
\end{equation}
For $\phi=\pi$ and $K=e^{-i\pi k}\in \mathbf{SU}(2)^{\otimes n}$ 
one finds $H\mapsto - H$ 
so $e^{-it H}$ is locally inverted.
\hfill$\blacksquare$

\vspace{3mm}
\noindent
However, for obtaining a both necessary {\em and} sufficient condition, it seems one has to sacrifice
the illustrative simplicity of geometry.

\begin{lemma}\label{lem:ad-inversion}
For a $U = e^{-it H}$ to be locally invertible at all $t$, it is both necessary and sufficient that
there is a local Hamiltonian $k$ in the orthocomplement $I^\bot_{\rm H}$ 
and a suitable $\phi\in[0,2\pi[$ with
\begin{equation}\label{eqn:ad-series}
    \sum\limits_{\ell=0}^\infty \frac{1}{\ell\; !} {\rm ad}^\ell_{\,(-i\phi\negthinspace k)} (H) 
		= - H\quad .
\end{equation}
\end{lemma}

\noindent
{\bf Proof:} Immediate consequence of the well-known identity
\begin{equation}
  e^X Y e^{-X} = \sum\limits_{\ell=0}^\infty \frac{1}{\ell\, !} {\rm ad}^{\,\ell}_X (Y) \quad ,
\end{equation}
where ${\rm ad}^{\,\ell}_X (Y)$ is the $\ell$-fold commutator of $X$ with $Y$, i.e. 
$[X,[X,\dots[X,Y]\cdots]]$ and
$\tfrac{1}{0!}{\rm ad}_X^{\,0} := \unity$.
Note this includes Lemma~2 as a special case.
\hfill$\blacksquare$

\subsection*{Algebra I: Eigenoperators}
\noindent
To begin with, 
recall the well-known fact that for two matrices $A,B$ to be similar i.e.
$XAX^{-1}=B$ with a non-singular $X$, their eigenvalues have to coincide.
Thus a Hamiltonian $H$ is invertible by unitary conjugation,
if and only if its non-zero eigenvalues (including multiplicity) all occur in pairs
of positive and negative sign. This is a necessary and sufficient condition
for inversion under some $U\in \mathbf{SU}(2^n)$, whereas for {\em local} unitary inversion 
by a $K\in \mathbf{SU}(2)^{\otimes n}$
being a special case it is merely a necessary one.

\vspace{3mm}
{\em Complete Basis for Locally Invertible Hamiltonians}\\[3mm]
\noindent
Although the decomposition of the algebra $\mathfrak{su}(2)^{\oplus n}$
into invariant subspace and orthocomplement
is illustrative, it is very tedious to be carried out  
case-by-case for each and every given Hamiltonian $H$. 
Rather, in order to obtain constructive parameters,
we will turn to the group $\mathbf{SU}(2)^{\otimes n}$ of local unitaries
and give a basis set to its eigenspace in which {\em all} locally
invertible Hamiltonians (of type-I) can be spanned.
To this end, observe that due to the series expansion,
\begin{eqnarray}
K e^{-it H} K^{-1} &=& e^{-it (K H K^{-1})} = e^{+it H} \quad\forall t\in\R{}\\
\label{U_inv}
\Leftrightarrow 
K H K^{-1} &=& \Ad K (H) = -H \quad , \label{H_inv}
\end{eqnarray} 
where the latter in turn is equivalent to the series expansion of $\Ad X$ in Lemma~\ref{lem:ad-inversion} and
Eqn~\ref{eqn:ad-series}.
Moreover, by the Kronecker product and the notation of a matrix as a vector (\/`vec\/')
consisting of the matrix columns stacked one upon another \cite{HJ1,HJ2}
one has illustrated the following obvious necessary and sufficient criterion for local
invertibility given as assertion (1)in the following

\noindent
\begin{lemma}
(1) The propagator $e^{-it H}$ is locally invertible for all $t\in\R{}$ if and only if
${\rm vec}\,H$ is
eigenvector of $\Ad K$ (here represented as $K^*\,\otimes K$)
to the eigenvalue $-1$:
\begin{equation}
(K^*\otimes K) \; {\rm vec}\, H = - {\rm vec}\, H
\quad .
\end{equation}
(2) The eigenspace to the eigenvalue $-1$ spans all the locally invertible
Hamiltonians, and it can be given in closed algebraic form by recursively making use of the eigenvectors
in $\mathbf{SU}(2)$, as $K \in \mathbf{SU}(2)^{\otimes n}$. 
\end{lemma}

\noindent
{\bf Proof:} for assertion (2), we give a constructive proof in view of explicit applications.
\vspace{2mm}

\noindent
{\bf Eigenvectors of the $\Ad K$-Superoperator} $K^*\,\otimes K$:\\
\noindent
For any local unitary $K \in \mathbf{SU}(2)^{\otimes n}$, the superoperator
$K^*\,\otimes K$ is just a $2n$-fold tensor product of unitary
$2\times 2$ matrices.
Using the quaternion parameterisation
\begin{equation}
U := \cos\tfrac{\beta}{2} \unity - i \sin\tfrac{\beta}{2}(n_x\sigma_x + n_y\sigma_y + n_z\sigma_z)
\in \mathbf{SU}(2) 
\end{equation}
with $\sum_{\nu=x,y,z} n_\nu^2 =1$,
the eigenvalues 
$
\lambda_\pm = e^{\pm i \tfrac{\beta}{2}}
$
(let $\beta\neq 0$)
are associated with the orthonormal
	eigenvectors
\begin{eqnarray}
v_+ &:=& \frac{1}{\sqrt{2(1+n_z)}}\begin{pmatrix}-n_x+in_y\\1+n_z\end{pmatrix}\\
v_- &:=& \frac{1}{\sqrt{2(1+n_z)}}\begin{pmatrix}1+n_z\\n_x+in_y\end{pmatrix}
\quad ,
\end{eqnarray}
	where the limit $n_z\to -1$ is uncritical:
	one finds $v_+=\left(\protect\begin{smallmatrix}1\\0\protect\end{smallmatrix}\right)$ 
	and $v_-=\left(\protect\begin{smallmatrix}0\\1\protect\end{smallmatrix}\right)$.

\vspace{3mm}
\noindent
{\em One Spin-$\tfrac{1}{2}$ Qubit}\\

The $\Ad K$-superoperator $(K^*\,\otimes K)$ for a single spin qubit
thus shows the four eigenvalues $\lambda^*_\pm \lambda^{\phantom{*}}_\pm
=e^{\mp i\frac{\beta}{2}}e^{\pm i\frac{\beta}{2}}$ (being either 1 or $e^{\pm i\beta}$)
associated with the four orthogonal eigenvectors $v^*_\pm \otimes v^{\phantom{*}}_\pm$.
Consequently the eigenspace to the overall eigenvalue $e^{\pm i\beta}=-1$ is spanned by the basis set
\begin{equation}
E^{(-)} := \{v^*_-\otimes v^{\phantom{*}}_+, v^*_+\otimes v^{\phantom{*}}_-\}
\end{equation}
while the eigenbasis to the overall eigenvalue $+1$ reads
\begin{equation}
E^{(+)} := \{v^*_+\otimes v^{\phantom{*}}_+, v^*_-\otimes v^{\phantom{*}}_-\}
\quad .
\end{equation}

\begin{remark}
For fixed parameters $(n_x,n_y,n_z)$ one finds $E^{(-)}\;\bot\;E^{(+)}$.
Note, however, that every element in $\mathfrak{su}(2)$ can be spanned in both $E^{(-)}$ and 
$E^{(+)}$: 
e.g., vec($\sigma_z$) may be expanded in $E^{(-)}$ by $(n_x,n_y,n_z)=(\cos\theta,\sin\theta,0)$,
while in $E^{(+)}$ the expansion requires $(n_x,n_y,n_z)=(0,0,1)$. This re-expresses the trivial fact
that $\sigma_z$ is inverted by any $\pi$ rotation about some axis in the $xy$-plane, whereas it is 
invariant under $z$~rotation.
\end{remark}

For completeness, in the limit $\beta\to 0$ define 
$E^{(0)}:= \frac{1}{\sqrt{2}}\big\{{\rm vec}\, \sigma_x,{\rm vec}\, \sigma_y,{\rm vec}\, \sigma_z\big\}$.

\vspace{3mm}
\noindent
{\em Two Spin-$\tfrac{1}{2}$ Qubits}\\[2mm]
For two qubits, the eigenbasis to the overall eigenvalue 
$\lambda^*_{1\pm}\lambda^*_{2\pm}\lambda^{\phantom{*}}_{1\pm}\lambda^{\phantom{*}}_{2\pm}=-1$ 
consists of vectors of the following subtypes:\\

\noindent
{\bf Subtype 0}
Embedding of the two limiting $1$-spin cases 
with $|\beta_1|=\pi$ {\em or} $|\beta_2|=\pi$\\[2mm]
$E_{\pi,\unity} := \big\{E_1^{(-)}\otimes {\rm vec}(\unity)_2^{\phantom{-}}\big\} \cup 
	\big\{{\rm vec}(\unity)_1^{\phantom{-}}\otimes E_2^{(-)}\big\}$\\

\noindent
{\bf Subtype 1}
Inversion of one spin {\em or} the other spin
with $|\beta_1|=\pi, \beta_2=0$ or $\beta_1=0,|\beta_2|=\pi$\\[2mm]
$E_{\pi,0} := \big\{E_1^{(-)}\otimes E_2^{(0)}\big\} \cup \big\{E_1^{(0)}\otimes E_2^{(-)}\big\}$\\

\noindent
{\bf Subtype 2}
Rotation on both spins with $|\beta_1|+|\beta_2|=\pi(\mod 2\pi)$ and $\beta_1,\beta_2\neq 0$\\[2mm]
$E_{\beta_1,\beta_2} := \big\{E_1^{(-)}\otimes E_2^{(-)}\big\}$\\

\noindent
{\bf Subtype 3}
Rotation on one spin, commutation with the other spin
where $|\beta_1|=\pi, \beta_2\neq 0$ arbitrary, {\em or} $|\beta_2|=\pi, \beta_1\neq 0$ arbitrary\\[2mm]
$E_{\pi,\beta^\parallel} := \big\{E_1^{(-)}\otimes E_2^{(+)}\big\} \cup \big\{E_1^{(+)}\otimes E_2^{(-)}\big\}$\\

\noindent
{\em $n$ Spin-$\tfrac{1}{2}$ Qubits}\\[2mm]
The generalisation to $n$ qubits with
\begin{equation}
\lambda^*_{1\pm}\lambda^{\phantom{*}}_{1\pm}\lambda^*_{2\pm}\lambda^{\phantom{*}}_{2\pm}\;
\cdots\;\lambda^*_{\ell\pm}\lambda^{\phantom{*}}_{\ell\pm}\;\cdots\;%
					\lambda^*_{n\pm}\lambda^{\phantom{*}}_{n\pm}=-1
\end{equation}
is obvious, because the construction
follows the pattern described by the indices to the eigenspaces. One may go from $n-1$ spins to
$n$ spins by adding the $n$th index from the set $\{\unity,0,\beta,\beta_\parallel\}$
to each of the previous $n-1$-spin cases according to the subtype of embedding.
Subtype 0 means expand $E_{n-1}$ to $E_{n-1}\otimes{\rm vec \unity}$;
subtype 1 gives $E_{n-1}\otimes E^{(0)}$;
subtype 2 leads to $E_{n-1}\otimes E^{(-)}$;
subtype 3 results in  $E_{n-1}\otimes E^{(+)}$.

In view of constructive results, the above subtypes have been
expressed in terms of sets of consistent rotation parameters $(n_x,n_y,n_z)_\ell$
and rotation angles $\beta_\ell$ on every spin $\ell$.
A locally invertible Hamiltonian has to be expandible in at least one set of
these self-consistent parameter sets. 
$\hfill\blacksquare$

In larger spin qubit systems, these checks may become increasingly
tedious. However, physical problems are often confined to special
settings: a Hamiltonian may be constituted by pair interactions, or
in other instances, a Hamiltonian may be made up of terms that can
be grouped in combinations of interactions transforming like spherical tensors
of various $p$-quantum order. For these two practically
relevant cases, we present more convenient methods.

\subsection*{Ising and Heisenberg Pair Interactions}
\begin{table}[Ht!]
\begin{center}
\caption{Type-I Invertibility of Elementary Pair Interactions}\label{tab:pair-inter}
\begin{tabular}{cllclcc}
\hline\hline\\[-1mm]
Pair &\phantom{XX}& Expansion &\phantom{X}& Type-I Local &\phantom{}& Symmetry\\[0mm]
Interaction&& in Subtype&&Inversion by &&Class\\[2mm]
\hline\hline\\[-1mm]
ZZ && $E_{\pi,0}$ && $\pi(\perp 1)$ or: $\pi(1\perp)$ &&\op \\[2mm]
\hline\\[-1mm]
XX && $E_{\pi,0}$ && $\pi(z1)$ or: $\pi(1z)$ &&\op \\[1mm]
   && $E_{\beta_1,\beta_2}$ && $\beta_1(z1)-\beta_2(1z)$ &&\op \\[1mm]
   &&  && [e.g.: $\tfrac{\pi}{2}(z1-1z)$]&&\op \\[1mm]
   && $E_{\pi,\beta^\parallel}$ &&  $\pi (\perp 1) - \pi(1\dashv)$ &&\op \\[2mm]
\hline\\[-1mm]
XY && $E_{\pi,0}$ && $\pi(z1)$ or: $\pi(1z)$ &&\op \\[1mm]
   && $E_{\pi,\beta^\parallel}$ &&  $\pi (x1) \pm \pi(1y)$ &&\op \\[1mm]
   &&                           &&  or: $\pi (y1) \pm \pi(1x)$ && \\[2mm]
\hline\\[-1mm]
X(-X) && $E_{\pi,0}$ && $\pi(z1)$ or: $\pi(1z)$ &&\op \\[1mm]
      && $E_{\beta_1,\beta_2}$ && $\beta_1(z1)+\beta_2(1z)$ &&\op \\[1mm]
      &&  && [esp.: $\tfrac{\pi}{2}(z1+1z)$]&&\pp \\[1mm]
      && $E_{\pi,\beta^\parallel}$ &&  $\pi (\perp 1) + \pi(1\vdash)$ &&\op \\[2mm]
\hline\\[-1mm]
XXX && none &&--&&-- \\[1mm]
XXY && none &&--&&-- \\[1mm]
XYZ && none &&--&&-- \\[2mm]
\hline\hline
\end{tabular}
\end{center}
\end{table}
The pair interactions of Ising and Heisenberg type can easily be related to the
$-1$ eigenspaces as summerised in Tab.~\ref{tab:pair-inter}:
while the Ising-$ZZ$ interaction can only be expanded in the eigenspaces of Subtype 1, i.e. $E_{\pi,0}$,
Heisenberg-$XX$ and $XY$ interactions allow for expansions in Subtype 1 as
well as Subtype 2 ($E_{\beta_1,\beta_2}$). 

For brevity, in the table we use the short-hand notation
$(z1)$ for $\tfrac{1}{2}(\sigma_z\otimes\unity)$,
and 
$(z1\pm 1z$) for $\tfrac{1}{2}(\sigma_z\otimes\unity \pm \unity\otimes\sigma_z)$.
as well as
$(\perp 1)$ for $\tfrac{1}{2}(\sigma_x \cos\phi + \sigma_y\sin\phi)\otimes\unity$
and analogously with reference to some fixed $\phi$ 
we write
$(\dashv)$ and $(\vdash)$ for $\tfrac{1}{2}(\sigma_x \cos(\phi\pm\tfrac{\pi}{2}) 
		+ \sigma_y\sin(\phi\pm\tfrac{\pi}{2}))$.
For example, the Heisenberg $XX$ interaction can of course be inverted by $\pi$ $z$"~pulses on one or
the other qubit, but also by an antisymmetric $z$"~rotation on both qubits,
where the rotation angle is $\beta$ on qubit 1 and $\beta-\pi$ on qubit 2.
Note that inverting generic $ZZ$, $XX$, and $XY$ interactions requires
pulses that are {\em non-symmetric} with regard to permuting
qubits 1 and 2. In view of convenient extensions to networks of pair 
interactions, we write \;$\op$\qquad\; for a pair interaction of two qubits
that is inverted by such non-symmetric local pulses.
The only exception of different symmetry is the Heisenberg $X(-X)$ interaction, since
it can also be inverted by a {\em permutation symmetric} $\tfrac{\pi}{2}$
$z$"~pulse on both of the qubits expressed by \;$\pp$\qquad\;.

Note that none of the Heisenberg $XXX$ or $XXY$ or $XYZ$ interactions
is type-I invertible by local unitaries, because their interaction
Hamiltonians already fail the simple necessary condition of being invertible over
the entire unitary group: their non-zero eigenvalues 
do not occur in pairs of opposite sign. For instance, the eigenvalues
to the $XYZ$ interaction Hamiltonian
$H_{XYZ}:= \alpha (\sigma_x\otimes\sigma_x) + \beta(\sigma_y\otimes\sigma_y) + \gamma(\sigma_z\otimes\sigma_z)$
read
\begin{equation}
\begin{split}
\lambda_1 &= +\alpha + \beta - \gamma\\
\lambda_2 &= -\alpha + \beta + \gamma\\
\lambda_3 &= +\alpha - \beta + \gamma\\
\lambda_4 &= -\alpha - \beta - \gamma
\end{split}
\end{equation}
with $\alpha,\beta,\gamma\in\R{}$.
Clearly, unless at least one of the parameters $\{\alpha,\beta,\gamma\}$ vanishes,
there are no pairs of opposite sign thus limiting the type-I invertible interactions
to $ZZ$ or $XX$ or $XY$ type.

\subsection*{Coupling Graphs for Networks of Pair Interactions}

Coupling networks made up by pair interactions between qubits  
can conveniently be represented by graphs: each vertex denotes a qubit, 
and an edge connecting two qubits $k$ and $l$
then corresponds to a non vanishing pair interaction or coupling $J_{kl}$.
These may take the form of any of Ising or Heisenberg type interactions
described before.
We will discuss connected graphs that do not necessarily have
to be complete. 

As will be seen next, interactions with coupling topologies of bipartite graphs
have special properties.

\begin{figure}[Ht!]\label{fig:bipartite}
\begin{minipage}[t]{\textwidth}
\begin{center}
\setlength{\unitlength}{1mm}
\begin{picture}(50,50)(25,-25)
\thicklines
\matrixput(8.7,2.8)(10,0){1}(0,5){4}{\circle*{2.3}}
\matrixput(0,0)(10,0){1}(0,5){5}{\circle{2.2}}
\matrixput(1.2,0.1)(5,2){1}(0,5){4}{\line(5,2){7}}
\matrixput(1.2,0.1)(1,2){1}(0,5){2}{\line(1,2){6.3}}
\matrixput(1.2,5.1)(5,-2){1}(0,5){4}{\line(5,-2){7}}
\matrixput(8.7,-7.6)(10,0){1}(0,-5){2}{\circle*{2.3}}
\matrixput(0,-10)(10,0){1}(0,-5){2}{\circle{2.2}}
\matrixput(1.2,-15)(5,2){1}(0,5){2}{\line(5,2){7}}
\matrixput(1.2,-10)(5,-2){1}(0,5){1}{\line(5,-2){7}}
\dashline[+50]{3}(1.2,-10)(7.8,2.9)
\dottedline(0,-8.2)(0,-1)
\dottedline(8.7,-6.7)(8.7,2)
%
\end{picture}\\
\end{center}
\end{minipage}
\caption{In a bipartite coupling graph, only vertices of different colour $\circ$ or $\bullet$
are pairwise connected.}
\end{figure}
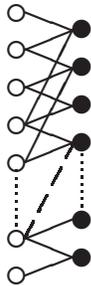

\vspace{3mm}
\noindent
\begin{lemma}[Variant to Beth \cite{WJB02}]
The evolution under Ising $ZZ$"~interactions
\begin{equation}
	H_{ZZ}:=\pi\sum\limits_{k<l}^n J_{kl}\; \tfrac{1}{2} \sigma_{kz}\otimes\sigma_{lz}
\end{equation}
is type-I invertible by local unitary operations if and only if its coupling topology
of non-vanishing couplings $J_{kl}$ forms a bipartite graph.
\end{lemma}
\noindent
{\bf Proof:}\\
{\em (i)}
For $H_{ZZ} \mapsto -H_{ZZ}$ it is sufficient that each of the edges of
the coupling graph is inverted. Using local actions on the vertices, this means 
every vertex of either the type $\bullet$ (or $\circ$) has to be inverted an odd number of times,
while the other type $\circ$ (or $\bullet$) remains invariant {\em i.e.} is inverted an even
number of times incl. zero.\\
{\em (ii)} 
Not only is this condition sufficient, it is also necessary: assume there were 
edges connecting two vertices of the same type (either $\bullet$ or $\circ$). Then
the couplings depicted by such edges would not be inverted, as they flip
their signs twice (or an even number of times) thus remaining effectively invariant. 
\hfill$\blacksquare$

\vspace{3mm}
\noindent
\begin{lemma}
The evolution under the  Heisenberg $XY$"~interaction
\begin{equation}
	H_{XY}:=\pi\sum\limits_{k<l}^n J_{kl}\; \tfrac{1}{2} 
		\big(\sigma_{kx}\otimes\sigma_{lx} + \kappa\;\sigma_{ky}\otimes\sigma_{ly}\big)\;,
\end{equation}
where $\kappa \in \;[-1;+1]$
is type-I invertible by local unitary operations if and only if (i) either its topology
of non-vanishing couplings $J_{kl}$ forms a bipartite graph or (ii)
$\kappa = -1$, in which case the coupling topology may take the form of any connected graph.
\end{lemma}
\noindent
{\bf Proof:}\\
Let $F_\nu$ with $\nu\in\{x,y,z\}$ denote the sum
over $n$ qubits with the Pauli matrix $\sigma_\nu^{(\ell)}$ in the $\ell^{\rm th}$ place
\begin{equation}
F_\nu:= \tfrac{1}{2}\sum\limits_{\ell=1}^n \unity^{(1)}\otimes\unity^{(2)}\otimes\cdots\otimes%
	\unity^{(\ell-1)}\otimes\sigma_\nu^{(\ell)}\otimes\unity^{(\ell+1)}\otimes\cdots\otimes\unity^{(n)}
\end{equation}
and analogously write $F_z^{(\circ)}$ or $F_z^{(\bullet)}$ if the sum just extends over
all qubits coloured $\circ$ or $\bullet$, respectively.\\
{\em (i)} For $\kappa \in \;]-1;+1]$ a bipartite coupling topology suffices to allow for
the inversion $H_{XY} \mapsto -H_{XY}$
by the rotations $\pi F_z^{(\circ)}$ or $\pi F_z^{(\bullet)}$.
A bipartite topology is also necessary, since in general no permutation-symmetric inversion of $H_{\rm XY}$
exists (see Tab.~\ref{tab:pair-inter}).\\
{\em (ii)} Cleary, also in the special case $\kappa = -1$ a bipartite coupling graph suffices.
However, it is not necessary, because
a $\tfrac{\pi}{2}$ $z$"~rotation on all the qubits 
($\tfrac{\pi}{2}\;F_z$) is invariant under qubit permutation and thus
does the same job on any connected coupling graph 
without requiring the distinction of a bipartite topology ({\em cp} the permutation symmetric
inversion of the $X(-X)$ interaction in Tab.~\ref{tab:pair-inter}).
\hfill$\blacksquare$

\subsection*{Examples of Pair Interaction Hamiltonians}
\noindent
For instance, neither Ising ZZ-coupling nor the Heisenberg XX and XY interactions on a
cyclic three-qubit coupling topology ($C_3$) are type-I invertible, because 
$C_3$ is clearly not bipartite. However, also on $C_3$, the Heisenberg X(-X)-interaction 
is type-I invertible, as will be illustrated below in the section on gradient flows.

\subsection*{Extension to Effective Multi-Qubit Interactions}

In multi-qubit effective interaction Hamiltonians on a coupling graph $G$, 
the interaction order $\ell$
(e.g. $\ell=3$ for $H_{\rm eff}=\sigma_z\otimes\sigma_z\otimes\sigma_z/2$)
may be used to group terms of different order. To each order, there is a
subgraph $G_{m_\ell}$.

\vspace{3mm}
\noindent
\begin{lemma}
Let $H_{\rm eff}$ be an effective multi-qubit interaction Hamiltonian
constituted by the $\ell$-interaction terms on the $m_\ell$ subgraphs.
\begin{equation}
        H_{\rm eff}=\sum\limits_{\ell,m_\ell} H_{\ell,{m_\ell}}
\end{equation}
where $\ell$ runs over the interaction orders and $m_\ell$ comprises
all the $\ell$-order interaction terms on subgraphs $G_{m_\ell}$.

Then $H_{\rm eff}$ is locally invertible of type-I if and only if
its constituents on the $G_{m_\ell}$ are all simultaneous eigenoperators of some $\Adr_K$
to the eigenvalue $-1$.
\end{lemma}

\noindent
{\bf Proof:}\\
The interaction order as well as the assignment to the subgraphs is $\Adr_K$ invariant.
\hfill$\blacksquare$

In simpler cases, the grouping may help to find local inversions by paper and pen.
However, more complicated multi-qubit interactions can be treated by exploiting the
transformation properties in terms of spherical tensors, as will be shown next.

\subsection*{Sequences of Interaction Propagators}
Clearly, a {\em palindromic} sequence of propagators 
\begin{equation}
U_s U_{s-1} \cdots U_2 U_1 U_1 U_2 \cdots U_{s-1} U_s 
\end{equation}
is locally invertible, if either
each component $U_k$ or at least one partitioning of the sequence is locally invertible.

\noindent
\subsection*{Multi-Qubit Interaction Hamiltonians of $p$-Quantum Order}

As usual in the treatment of angular momenta in spin"~$j$ representation
(where $j=n\,\cdot\,j'$ may sum the spin quantum numbers of $n$ identical, 
i.e. permutation symmetric single spins"~$j'$ to the group spin"~$j$)
one defines a rank"~$j$ spherical tensor $\TE jm$ of order $m$ by the transformation properties
under rotation by the Euler angles $\{\alpha,\beta,\gamma\}$
\begin{equation}
\begin{split}
\WigDabc j \,\,&\TE jm \,\,{\WigDabc{j}}^{-1} \\
	&= \sum_{m'=-j}^j\,\WigDabcm{m'}{m}{j}\, \TE{j}{m'}\,,
\end{split}
\end{equation}
where the elements
\begin{equation}
\begin{split}
\WigD{m'}{m}{j}(\alpha,\beta,\gamma) :&= %
\braket{j,\,m'}%
        {e^{-i\frac{\alpha}{2}\,\sigma_z^{(j)}}%
        e^{-i\frac{\beta}{2}\,\sigma_y^{(j)}}%
        e^{-i\frac{\gamma}{2}\,\sigma_z^{(j)}}\,\,j,\,m}\\
	&= e^{-im'\alpha}\,\,\,\Wigd{m'}{m}{j}(\beta)\,\,e^{-im\gamma}
\end{split}
\end{equation}
constitute the full Wigner rotation matrix. An equivalent definition of the spherical tensors
via the Pauli matrices or angular momentum operators $\{J_x, J_y, J_z\} \iso i\; \mathfrak{su}(2)$ 
in spin"~$j$ representation uses the commutation relations
\begin{equation}\label{eqn:tensor}
\begin{split}
\comm{J_x \pm i J_y}{\TE jm}\,&\equiv\comm{J^\pm }{\TE jm}\,\\&= \,\sqrt{j(j+1)-m(m\pm 1)}\,\,\,\TE j{m\pm 1} \\
\comm{J_z}{\TE jm}\,&=\,m\,\TE jm 
\end{split}
\end{equation}
and establishes the relation to the algebra $\mathfrak{sl}(2,\C{})$ represented by
$\{J_+, J_-, J_z\}$, as will be further illustrated below.

Now specialise $D^{(j)}(\alpha,\beta,\gamma)$ to $\WigDOOp j$ by setting
$\alpha = \beta = 0$ and $\gamma \equiv \phi$.
Moreover, identify $m = m'$ with the quantum order $p$ of the interaction 
Hamiltonian $H \equiv \TE jp$ to obtain the eigenoperator equation
\begin{equation}
\WigDOOp j \;\;\TE jp \;\;{\WigDOOp{j}}^{-1} = e^{-ip\phi}\; \TE{j}{p}\,.
\end{equation}

\subsubsection*{Inversion by Joint Local $z$-Rotations}
Note that the commutation relations for $\mathfrak{su}(2)$ or $\mathfrak{sl}(2,\C{})$ 
do not change, if one replaces
the $J_\nu$ where $\nu\in\{x,y,z;+,-\}$ by the symmetric sum
over $n$ qubits with $J_\nu^{(\ell)}$ in the $\ell^{\rm th}$ place written again as
\begin{equation}
F_\nu:= \sum\limits_{\ell=1}^n \unity^{(1)}\otimes\unity^{(2)}\otimes\cdots\otimes\unity^{(\ell-1)}\otimes
	J_\nu^{(\ell)}\otimes\unity^{(\ell+1)}\otimes\cdots\otimes\unity^{(n)}\,.
\end{equation}

Therefore, we may also envisage $\WigDOOp j$ as a {\em local} $z$"~rotation by an angle $\phi$
acting {\em jointly} on all the $n$ qubits, {\em i.e.}, $\WigDOOp j =: K(\phi,F_z) \in \mathbf{SU}(2)^{\otimes n}$.
Thus one arrives at the two identical formulations
\begin{equation}\label{eqn:tensor-AdK}
\begin{split}
K(\phi,F_z) \;\;\TE jp \;\;{K(\phi,F_z)}^{-1} &= e^{-ip\phi}\; \TE jp\\
 \Adr_{K(\phi,F_z)} \TE jp \,\,  &= e^{-ip\phi}\; \TE jp\;.
\end{split}
\end{equation}
Clearly, inverting $\TE jp$ to $-\TE jp$ 
requires $e^{-ip\phi} =-1$. By the
Fourier duality between the quantum order $p$ and the phase $\phi$ one readily
finds the results given in Table~\ref{tab:Fz-inversion}.
A joint local $z$"~rotation of angle 
$\pi/r$ (with a fixed $r=1,2,3, \dots$) {\em simultaneously} inverts 
all the rank-$j$ tensors of different quantum orders $p$
given in the same row of the table. Thus also any linear combination
of interaction tensors of quantum orders 
$\pm p=r(2q+1)$ can be inverted at a time for all $q=1,2,3,\; \dots$\;.
\begin{table}[Ht!]
\begin{center}
\caption{Inversion of Spherical Tensors by Joint $z$-Rotations}\label{tab:Fz-inversion}\label{tab:fourier}
\begin{tabular}{ccc}
\hline\hline\\[-3mm]
rotation angle $\phi$ && inverts interactions of quantum order $\pm p$\\
\hline\\[-3mm]
${\pi}$ && $1,3,5,\cdots,2q +1\;\leq\;j$\\
${\pi}/{2}$ && $2,6,10,\cdots,4q + 2\;\leq\;j$\\
${\pi}/{3}$ && $3,9,15,\cdots,6q+3\;\leq\;j$\\
$\vdots$ && $\vdots$\\
${\pi}/{r}$ && $r,3r,5r,\cdots,r\,(2q + 1)\;\leq\;j$\\
\hline\hline
\end{tabular}\hspace{15mm}
\end{center}
\end{table}

Obviously, interaction Hamiltonians represented by spherical tensors $\TE j0$
of order $p=0$ cannot be inverted by $z$"~rotations $K(\phi,F_z)$, but may possibly be 
inverted by local unitaries operating on other than the $z$"~axes. However, rank"~$0$
tensors $\TE 00$ transforming like pseudoscalars such as e.g.\ the Heisenberg $XXX$
interaction Hamiltonian, which in two spins is proportional to
\begin{equation}
 \TE 00 = \frac{-1}{2 \sqrt{3}}\;\big(\sigma_x\otimes\sigma_x + \sigma_y\otimes\sigma_y + \sigma_z\otimes\sigma_z\big)\;,
\end{equation}
cannot be inverted at all. Not only does this hold for local unitaries, but for
any similarity transform by some $X\in\mathbf{GL}(2^n)$, since the 
non-zero eigenvalues of \TE 00 do not occur in pairs
of opposite sign ({\em vide supra}).

\subsubsection*{Inversion under Individual Local $z$-Rotations}
Since in a single spin qubit, $\Ad {\phi,z}$ has the eigenoperators
$J_\nu \in \{\unity, J_z, J_+, J_-\}$ associated to the respective eigenvalues 
$e^{-i p_\nu \phi} \in \{1,1,e^{-p_+\phi},e^{-p_-\phi}\}$,
it is easy to generalise the previous arguments to the case of $z$"~rotations
on $n$ qubits---but with individually differing rotation angles on each
spin qubit $\phi_1, \phi_2, \dots, \phi_\ell, \dots, \phi_n$.
Now consider a Hamiltonian $\bar H$ taking the special form of a tensor
product of $\Ad {\phi,z}$ eigenoperators on each spin qubit $\ell=1,\dots, n$
according to
\begin{equation}
\bar H := \JZ 1 \otimes \JZ 2 \otimes \cdots \otimes \JZ \ell \otimes \cdots \otimes \JZ n
\end{equation}
with independent $\nu_\ell \in \{\unity, z, +, -\}$ 
on each spin qubit. Then $\bar H$ is clearly an eigenoperator to
individual local $z$-rotations $K(\phi_1,\; \dots\;, \phi_n, F_z) \in \mathbf{SU}(2)^{\otimes n}$ 
by virtue of
\begin{equation}
\Ad {K(\phi_1,\; \dots\;, \phi_n, F_z)} \big(\bar H\big) = %
	e^{-i (p_1 \phi_1 +\; \cdots \;+ p_n \phi_n)} \; \bar H\;.
\end{equation}
So $\bar H$ is inverted if there is a set of rotation angles $\{\phi_\ell\}$
with
\begin{equation}
\sum\limits_{\ell = 1}^n p_\ell \phi_\ell = \pm \pi\; (\mod 2 \pi)\quad ,
\end{equation}
which is the case if there is at least one spin qubit $\ell$ giving rise to an
interaction of quantum order $p_\ell = \pm 1$. Moreover, a linear
combination of such Hamiltonians 
$\bar H_\Sigma := \sum\limits_{\lambda=1}^m c_\lambda \bar H_\lambda$ 
is jointly invertible by an individual local $z$"~rotation 
$K(\phi_1,\; \dots\;, \phi_n, F_z)$ ,
if there is at least one consistent
set of rotation angles $\{\phi_\ell\}$ simultaneously satisfying
for all the components $\bar H_\lambda$
\begin{equation*}
\sum\limits_{\ell = 1}^n p_{\lambda,\ell}\;\cdot\; \phi_\ell = \pm \pi\; (\mod 2 \pi)\;,
\end{equation*}
which expresses the linear system
\begin{equation}\label{eqn:linsys}
\begin{pmatrix} 
	p_{11} & p_{12} & \cdots & p_{1n} \\
	p_{21} & p_{22} & \cdots & p_{2n} \\
	p_{31} & p_{32} & \cdots & p_{3n} \\
	\vdots & \vdots & \ddots & \vdots \\
	p_{m1} & p_{m2} & \cdots & p_{mn} \\
\end{pmatrix} 
\begin{pmatrix}\phi_1\\ \phi_2\\ \vdots\\ \phi_n\end{pmatrix} =
\begin{pmatrix}\pm \pi\; (\mod 2 \pi)\\ \pm \pi\; (\mod 2 \pi)\\ \pm \pi\; (\mod 2 \pi)\\ \vdots\\ %
		\pm \pi\; (\mod 2 \pi)\end{pmatrix}\;.
\end{equation}
Note the signs on the {\sc rhs} may be chosen independently in $2^m$ ways
with every choice forming a system of $m$ linear equations in $n$ variables. 
Therefore,
if the vector of any of the combinations $\pi\;(\pm 1, \pm 1, \; \cdots \; , \pm 1)^t$
can be expanded in terms of the
column vectors of $P:=\big(p_{\lambda,\ell}\big)$ with real coefficients, 
then $\bar H_\Sigma$ is locally invertible by $z$"~rotations.
In the special case of $m=n$ and $P$ non-singular, there always is a consistent set of 
individual rotation angles for inverting $\bar H_\Sigma$ for any choice of signs.
For simplicity, we will drop the index $\Sigma$ in $\bar H_\Sigma$ henceforth
writing $\bar H$ for Hamiltonians locally invertible by individual $z$"~rotations.\\


\begin{corollary}
Let $\bar H$ be locally invertible by individual $z$"~rotations on each qubit. Then
the following hold.
\begin{itemize}
\item[(1)] Any Hamiltonian $H$ on the local unitary orbit $\Ad {\mathbf K} (\bar H)$ of any such
	   $\bar H$ generates a one-parameter unitary group that is locally invertible of type-I.
\item[(2)] In turn, any type-I locally invertible Hamiltonian $H$ is on a local
	   unitary orbit of some $\bar H$.
\end{itemize}
\end{corollary}

\noindent
{\bf Proof:}
(1) is obvious. (2) follows since every local unitary $K$ is locally unitarily similar to a local $z$"~rotation $K_z$:
$K = \bar K K_z \bar K^{-1} \Rightarrow
K H K^{-1} = - H \Leftrightarrow K_z \bar H K_z^{-1} = - \bar H$ where $ \bar H := \bar K^{-1} H \bar K$.
\hfill$\blacksquare$\\
Thus local invertibility by $z$"~rotations can be looked upon as the normal form of
the problem.

Moreover, in order to see the link to the root space decomposition of the
semisimple Lie algebra $\mathfrak{sl}(2,\C{})$ in spin"~$j$ representation,
take the derivative  of Eqn.~\ref{eqn:tensor-AdK}
\begin{eqnarray}
\Partial{}{\phi}\big|_{\phi=0}
\Adr_{K(\phi,z)} \TE jp \,\,  &=& \Partial{}{\phi}\Big|_{\phi=0} e^{-ip\phi}\; \TE jp\\[3mm]
\Leftrightarrow\;
-i \adr_{k_z}(\TE jp) &=& -ip\; \TE jp\;.
\end{eqnarray}
(NB: the minus sign on the left side of the last identity is due to 
the convention $K(\phi,z):= e^{-i\phi k_z}$ analogous to $U:=e^{-itH}$ imposed
by Schr{\"o}dinger's equation.)

Note the tensors are eigenoperators to $\adr_{k_z}$ of the joint local $z$"~rotations as anticipated
in Eqn.~\ref{eqn:tensor}.

\subsection*{Algebra II: Root-Space Decomposition}\label{sec:root_space}
To fix notations, let $\frak{g}$ be a complex semisimple
Lie algebra and let $\frak{g}_0$ be a Cartan subalgebra
of $\frak{g}$, i.e. a maximally abelian subalgebra such
that for all $H \in \frak{g}_0$, 
	\footnote{In accordance with the standard literature
	on Lie algebras, in this paragraph the letter $H$ denotes
	elements $H \in \frak{g}_0$ of the Cartan subalgebra, not Hamiltonians.}
the commutator superoperators $\adr_H$ are 
simultaneously diagonalisable.
Then the root-space decomposition
of $\frak{g}$ with respect to $\mathfrak g_0$ takes the form
\begin{equation}
\frak{g} = \frak{g}_0 \oplus \bigoplus_{\alpha\neq 0}\frak{g}_{\alpha}\quad, 
\end{equation}
where the root-spaces $\frak{g}_{\alpha}$ (with $\alpha \neq 0$) are the non-trivial
simultaneous eigenspaces 
\begin{equation}
\frak{g}_{\alpha} := \{g \in \frak{g}\, |\, \adr_H(g) = \alpha(H)\, g\,\}\quad .
\end{equation}
The corresponding non-trivial $\alpha$'s are called roots of the decomposition. 
They are elements of the dual space $\frak{g}_0^*$ of linear functionals on $\frak g_0$.

In the following, we consider the complex semisimple Lie algebra 
$\mathfrak{sl}(N,\C{})$ as the complexification
of the real Lie algebra $\mathfrak{su}(N)$.
Define $E_{ij}$ as a square matrix differing from the zero matrix by just
one element---the unity in the $j^{\rm th}$ column of the $i^{\rm th}$ row.
Moreover, let $\frak{g}_0$ be the
set of all diagonal matrices in $\frak{sl}(N,\C{})$ and define
$e_{i}(H) := H_{ii}$ for all $H \in \frak{g}_0$. Then for every
$H \in \frak{g}_0$, the $E_{ij}$ are {\em simultaneous} eigenoperators
of the commutator superoperators $\adr_H$ with eigenvalues depending linearly on $H$
\begin{equation}\label{eqn:eigen_ad}
\begin{split}
\adr_H(E_{ij}) &= (H_{ii}-H_{jj})\,E_{ij}\\
	 &= \big(e_i(H)-e_j(H)\big)\,E_{ij} =: \alpha_{ij}E_{ij}\quad.
\end{split}
\end{equation}
Thus the root-space decomposition of $\frak{sl}(N,\C{})$ may be rewritten as
\begin{equation}
\mathfrak g = \mathfrak g_0 \oplus\;\bigoplus\limits_{i\neq j} \C{} E_{ij}\quad.
\end{equation}
\begin{figure*}[Ht!]
\includegraphics[width=0.95\textwidth]{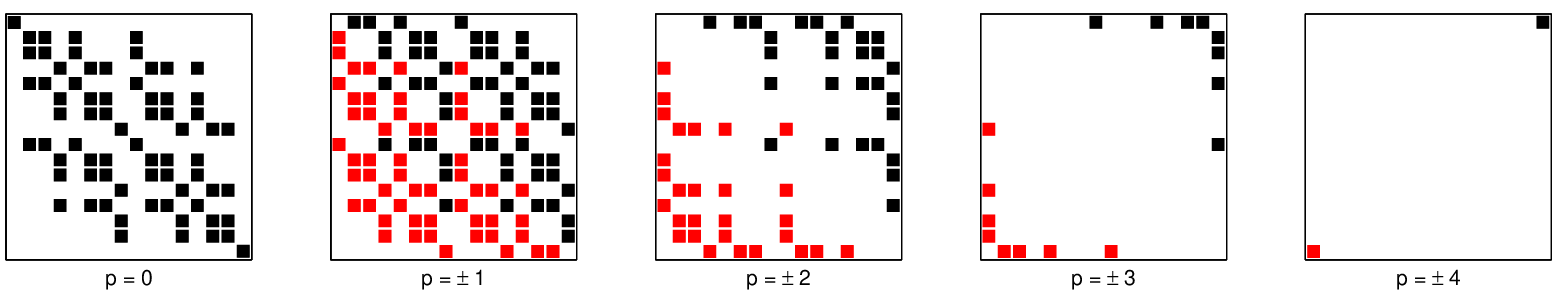}
\caption{(Colour online) 
Matrices $E_{ij}$ constituting the rank-$4$ spherical tensors $T_{4,p}$ of order $p$
(here in a system of $4$ spins-$\tfrac{1}{2}$, $p\in[-4,+4]$).
The non-zero elements in the $E_{ij}$ are marked $\blacksquare$ for 
$p\geq 0$ and \RED{$\blacksquare$} for $p < 0$.
The $T_{4,p}$ are eigenoperators to the Weyl torus element $F_z\in\mathfrak t$ according to 
$\adr_{F_z}(T_{4,p}) = p\; T_{4,p}$ with the eigenvalues being the quantum orders $p$.
The constituents of the tensors are the single-element Weyl matrices $E_{ij}$ sharing the same eigenvalues
$\adr{F_z}(E_{ij}) = p\; E_{ij}$.
According to Tab.~\ref{tab:fourier}, a Hamiltonian comprising elements that transform like
$T_{4,\pm p}$ can locally be sign-inverted e.g. by a joint $z$-rotation of angle $\pi/p$.
} 
\end{figure*}

Furthermore, if $\frak{g}'$ is a {\em compact} real semisimple Lie algebra
with maximally torus algebra $\frak{t}'$, then the complexification 
$\frak{t}$ of $\frak{t}'$ gives a Cartan subalgebra of the 
complexification $\frak{g}$ of $\frak{g}'$. For example in the case of $\frak{su}(N)$,
\begin{equation}
\frak{t}' := \{i\,\diag(\theta_1, \theta_2,\,\dots\,,\theta_N)\,|\, \sum_\ell\theta_\ell = 0\}
\end{equation}
may be chosen as maximal torus algebra. Then
\begin{equation}
\frak{t} := \frak{t'} + i\, \frak{t'} 
\end{equation}
i.e. the set of all complex diagonal matrices forms the Cartan subalgebra
of $\frak{sl}(N,\C{})$. 

Now Eqn.~\ref{eqn:eigen_ad} shows that an $E_{ij}$ with $i\neq j$ can be sign-inverted
provided $(H_{ii}-H_{jj})\neq 0$. The generic case, joint and individual local
$z$"~rotations are specified next.
\begin{proposition}
In a system of $n$ spins-$\tfrac{1}{2}$, 
for the single-element matrices $E_{ij}$ with $i\neq j$ the following hold:
\begin{itemize}
\item[(1)] to any $E_{ij}$ there is an element of the Weyl torus
		taking the form $T=\exp(-i \diag\big(\theta_1,\theta_2,\,\dots\,,\theta_N)\big)$
		so that $\Adr_T(E_{ij})= -E_{ij}$;
\item[(2)] any matrix $E_{ij}$ can also be sign-inverted by a single local $z$"~rotation;
\item[(3)] in contrast, by a joint local $z$"~rotation on all the $n$ spins-$\tfrac{1}{2}$,
		an $E_{ij}$ can only be sign-inverted if
		for its indices $i,j$ the reductions by $1$ written 
		as binary numbers $(i-1)_2$ and $(j-1)_2$
		do not have the same number of $0$'s and $1$'s 
		(irrespective of the order).
		
\end{itemize}
\end{proposition}
\noindent
{\bf Proof:}
First, note that although $\mathfrak{sl}(2^n,\C{})$ comprises the generators of
{\em all} special unitary propagators, its maximally abelian subalgebra $\mathfrak g_0$
can---without loss of generality---always be chosen such that it includes the
generators of {\em all the local} $z$"~rotations. They suffice to be considered,
since the $E_{ij}$ are simultaneous eigenoperators to all elements in $\mathfrak g_0$.\\[-2mm]

\noindent
(1) Obviously elements in the Weyl torus algebra $\mathfrak t$
can be chosen such that $\theta_i - \theta_j \neq 0$.\\[-2mm]

\noindent
(2) Clearly any off-diagonal element $E_{ij}$ in the boxes of the block
matrix $H:=\begin{pmatrix} \phantom{.}\unity & \Box \\ \Box & -\unity \end{pmatrix}$
is associated with a non-zero root $(e_i-e_j)(H) = H_{ii}-H_{jj}$. The 
same holds true for $H\otimes\unity$ and $\unity\otimes H$. 
Likewise any off-diagonal element $E_{ij}$ is in one of the boxes of the
following embedded Pauli $z$"~matrices
$$ \tilde\sigma_{\ell z} := \unity_2^{\otimes (\ell -1)}\otimes 
\begin{pmatrix}
\phantom{.}1 & \Box \\ \Box & -1
\end{pmatrix}_{(\ell)}
\otimes \unity_2^{\otimes(n-\ell)} \quad,$$
where the box sizes coincide with $\unity_2^{\otimes(n-\ell)}$ (set $\unity_2^{\otimes 0}=1$).
Let the index run $\ell=1,2,\,\dots\,n$ to see that in fact every off-diagonal element $E_{ij}$
can be associated with some $\ell$, which implies 
any $E_{ij}$ can be sign-inverted by at least one local $z$"~rotation on some single spin qubit $\ell$.
(Due to permutation symmetry, for $\ell<n$
there are off-diagonal $E_{ij}$ with non-zero roots even outside the boxes; they
add further options of choosing a qubit $\ell$.)\\[-2mm]

\noindent
(3) 
For $E_{ij}$, let $(i-1)_2=:\sum_{k=0}^{n-1} 2^k b_k$ and $(j-1)_2=:\sum_{k=0}^{n-1} 2^k b_k'$ 
define the $n$"~digit binary representations of the indices reduced by 1.
If $(i-1)_2$ and $(j-1)_2$ have the same number of $0$'s and $1$'s,
we will show that $E_{ij}$ belongs to a zero root of $F_z$, {\em i.e.} $\adr_{F_z}(E_{ij})=0$.
In Appendix"~B we gave a general formula for the matrix elements $(F_z)_{ii}$.
So
\begin{equation}\label{eqn:adFz0}
{(F_z)}_{ii} - {(F_z)}_{jj} = 
	\tfrac{1}{2}\;\sum_{k=0}^{n-1}(-1)^{b_k} -\tfrac{1}{2}\;\sum_{k=0}^{n-1}(-1)^{b'_k}
\end{equation}
vanishes, if and only if an equal number of terms $(-1)^0$ and $(-1)^1$ appears in both 
$n$"~term sums as claimed.  \hfill$\blacksquare$


\begin{corollary}
As the maximally abelian algebra $\mathfrak g_0$ of both $\mathfrak{sl}(2,\C{})$
and
$\mathfrak{su}(2)$ 
in the representation of $n$ spins"~$j'$ 
can always be chosen such as to comprise the
generators $k_z$ of
$K(\phi,F_z) \in \mathbf{SU}(2)^{\otimes n}$
bringing about local $z$"~rotations jointly on all spins,
the tensors of order $p$
are associated to the root space elements $E_{ij}$ of $\mathfrak{sl}(2,\C{})$ showing
the eigenvalue $(e_i-e_j)(F_z)=p$.
\end{corollary}

Allowing individual $z$"~rotations on each qubit again, one finds
\begin{equation}
\Ad {K(\phi_1,\; \dots\;, \phi_n, F_z)} \big(E_{ij}) = %
        e^{-i (p_1 \phi_1 +\; \cdots \;+ p_n \phi_n)} \; E_{ij}\;,
\end{equation}
because any $E_{ij}$ can be written as a tensor product of the single-element
two by two matrices 
$\{J^\alpha,J^\beta,J_+,J_-\}:=\left\{
	\left(\begin{smallmatrix}1 & 0\\ 0 & 0\end{smallmatrix}\right),\;
	\left(\begin{smallmatrix}0 & 0\\ 0 & 1\end{smallmatrix}\right),\;
	\left(\begin{smallmatrix}0 & 1\\ 0 & 0\end{smallmatrix}\right),\;
	\left(\begin{smallmatrix}0 & 0\\ 1 & 0\end{smallmatrix}\right)\;
\right\}$ 
associated
with the eigenvalues $e^{-ip_\nu\phi}$ for $\nu\in\{\alpha,\beta,+,-\}$,
where $p_\alpha = p_\beta = 0$ and $p_\pm = \pm 1$.
\vspace{2mm}

\noindent
{\bf Examples:}\\[1mm]
(1) The single-element Weyl matrix $E_{08,15}$ belongs to the zero-root $(e_i-e_j)(F_z^n)=p=0$
since for all $n\geq4$ the binary representations (according to Proposition 1.3) 
end with $7_2=0111$ and $14_2=1110$ having the same number of $0$'s and $1$'s.
Thus it cannot be sign-inverted by {\em joint} local $z$"~rotations, whereas
by being off-diagonal it can always be sign-inverted by an {\em individual} local $z$"~rotation.\\[1mm]
(2) In contrast, for $E_{47,11}$ one finds by Eqn.~\ref{eqn:adFz0} 
and the binaries
$46_2=101110$ and $10_2=001010$ that $p=(-4)\tfrac{1}{2}$.
So it can be sign-inverted by a joint local $z$"~rotation with rotation angle $\phi=\tfrac{\pi}{4}$
in accordance with Tab.~\ref{tab:fourier}.

\vspace{2mm}
Given the relation to the transformation properties of spherical tensors,
it is easy to analyse type"~I local invertibility of linear combinations of 
root space elements $E_{ij}$ under joint or individual $z$"~rotations.

\begin{proposition}
In a system of $n$ qubits,
a linear combination of single-element matrices $E_{ij}$  
\begin{equation}
E_\Sigma := \sum\limits_{\lambda = 1}^m c_\lambda E_{ij}^{(\lambda)}
\end{equation}
with $i\neq j$ and $c_\lambda\in\C{}$
is sign-invertible by an individual local $z$"~rotation
$K(\phi_1,\; \dots\;, \phi_n, F_z)$ ,
if there is at least one consistent
set of rotation angles $\{\phi_\ell\,|\,\ell=1,2,\dots,n\,\}$ simultaneously satisfying
for all its constituents $E_{ij}^{(\lambda)}$
\begin{equation*}
\sum\limits_{\ell = 1}^n p_{\lambda,\ell}\;\cdot\; \phi_\ell = \pm \pi\; (\mod 2 \pi)\;,
\end{equation*}
which coincides with the linear system in Eqn.~\ref{eqn:linsys}.
\end{proposition}

\subsection*{Relation to Time Reversal and Cartan Decompositions}

As will be shown, the detailed discussion of the root-space
decomposition in the previous section was in fact needed,
and a mere Cartan decomposition does not decide about type-I
local invertibility.

Let $\frak{g}$ be a real compact semisimple Lie algebra and
let the mapping $\theta: \frak{g} \to \frak{g}$ be any involutive (Lie algebra)
automorphism. Then $\theta$ defines a Cartan-like decomposition
$\frak{g} = \frak{k}_{\theta} \oplus \frak{p}_{\theta}$ of $\frak{g}$,
where $\frak{k}_{\theta}$ and $\frak{p}_{\theta}$ are
the respective $+1$ and $-1$ eigenspaces of $\theta$, i.e.
\begin{eqnarray}
\theta(X) &=& \phantom{-}X \quad \text{for all $X\in \mathfrak k_\theta$}\\
\theta(X) &=& -X \quad \text{for all $X\in \mathfrak{p_\theta}$} 
\end{eqnarray}
ensuring the standard commutation relations
\begin{eqnarray}
\protect{[{\mathfrak k_\theta}, {\mathfrak k_\theta}]} &\subseteq& {\mathfrak k_\theta}\\
\protect{[{\mathfrak k_\theta}, {\mathfrak p_\theta}]} &\subseteq& {\mathfrak p_\theta}\\
\protect{[{\mathfrak p_\theta}, {\mathfrak p_\theta}]} &\subseteq& {\mathfrak k_\theta} \quad .
\end{eqnarray}
In $\mathfrak{su}(2^n)$, one may choose 
the so-called concurrence Cartan involution \cite{BBO05}
\begin{equation}
\theta_{\rm CC}(X) := (-i\sigma_y)^{\otimes n} X^* \big((-i\sigma_y)^{\otimes n}\big)^\dagger\;,
\end{equation} 
where $\theta_{\rm CC}$ takes the form of the bit flip operator and thus relates to time reversal.
Bullock {\em et al.} ~\cite{BBO05} classified Hamiltonians $iH \in \mathfrak p_{\rm CC}$
as symmetric with respect to time-reversal and those in $\mathfrak k_{\rm CC}$ as anti-symmetric.
Since $(-i\sigma_y)^{\otimes n} = e^{-i\pi F_y}$, this representation of the Cartan
involution is equivalent to a local $y$"~rotation acting jointly on $n$ qubits 
following complex conjugation. Due to the latter, the Cartan involution $\theta$ is unphysical
(as also pointed out in ref.~\cite{BBO05}) and it is thus distinct from the local unitary operations 
discussed here.

Note that in $\mathfrak{su}(4)$ $\mathfrak k_{\rm CC}$ coincides with the
algebra $\mathfrak k$ generating the local unitaries $\mathbf K := \mathbf{SU}(2)^{\otimes 2}$,
which is the reason for our notation, whereas in $\mathfrak{su}(2^n)$ with $n\geq 3$ this
is no longer true, since $\mathfrak k_{\rm CC}$ comprises $m$"~linear interaction Hamiltonians
with $m$ odd, while $\mathfrak p_{\rm CC}$ encompasses those with $m$ even.

As described above for the simple case of two qubits, the pair interactions ($m=2$) 
$H_{\rm ZZ}, H_{\rm XY}\in \mathfrak p_{\rm CC}$ are locally type-I invertible, while
$H_{\rm XXX}, H_{\rm XYZ} \in \mathfrak p_{\rm CC}$ are not. 
Yet in two qubits $\mathfrak k_{\rm CC}=\mathfrak k$ (s.a.),
so all the elements in $\mathfrak k_{\rm CC}$ are---by definition---type-I invertible, while for $3$ and more
qubits, $\mathfrak k_{\rm CC}$ generically contains type"~I invertible interactions ({\em e.g.} $zzz$)
as well as  non-invertible ones ({\em e.g.} $H_{\rm xxx + 2 yyy + 3 zzz}$). 

Hence, a Cartan-type decomposition into time-reversal symmetric and antisymmetric subspaces
does not decide whether an interaction is locally invertible or not.
In $\mathbf{SU}(N)$ with $N:=2^n$,
also for other standard choices of the Cartan involution, such as \cite{Helgason78}
\begin{eqnarray}
&\theta_{\rm A\,I}(X) &:= X^*    \\[1mm]
&\theta_{\rm A\,II}(X) &:= J_N X^* J_N^{-1}  \\[1mm]
&\theta_{\rm A\,III}(X) &:= I_{p,q} X\, I_{p,q} 
\end{eqnarray} 
with $p=q=N/2$ in the definitions
\begin{eqnarray}
J_N &:=& \protect{\begin{pmatrix} 0 & \unity_{N/2} \\ -\unity_{N/2} & 0 \end{pmatrix}}  \\[1mm]
I_{p,q} &:=& \protect{\begin{pmatrix} \unity_p &  0 \\ 0 & -\unity_q \end{pmatrix}}
\end{eqnarray} 
the decomposition into
$\mathfrak k_\theta$ and $\mathfrak p_\theta$ does never completely agree with the subdivision
into locally invertible and non-invertible interacion Hamiltonians $i H_{\rm int}\in\mathfrak{su}(2^n)$
as shown in Tab.~\ref{tab:cart-inv}.
Rather in the general case,
it takes the more specific patterns derived from the root space decomposition
as described in the previous section.
\begin{table}[Ht!]
\begin{center}
\caption{Relation to Cartan Decomposition for Different Choices of Involution}\label{tab:cart-inv}
\begin{tabular}{llccccc}
\hline\hline\\[-1mm]
$H_{\rm interaction}$ &&Type-I & $\theta_{\rm CC}$ & $\theta_{\rm A\,I}$ & $\theta_{\rm A\,II}$ &\phantom{XX}$\theta_{\rm A\,III}$ \\
\multicolumn{2}{l}{\small(number of qubits $m$)}& invertible && && \\[2mm]
\hline\\[-1mm]
Pauli & $X1$ 	  & $+$ & $\mathfrak k$ & $\mathfrak p$ & $\mathfrak k$ & $\mathfrak p$\\
matrices & $Y1$ 	  & $+$ & $\mathfrak k$ & $\mathfrak k$ & $\mathfrak k$ & $\mathfrak p$\\
      & $Z1$ 	  & $+$ & $\mathfrak k$ & $\mathfrak p$ & $\mathfrak p$ & $\mathfrak k$\\[2mm]
$m=2$ & $H_{ZZ}$  & $+$ & $\mathfrak p$ & $\mathfrak p$ & $\mathfrak p$ & $\mathfrak k$\\
      & $H_{XX}$  & $+$ & $\mathfrak p$ & $\mathfrak p$ & $\mathfrak k$ & $\mathfrak p$\\
      & $H_{XY}$  & $+$ & $\mathfrak p$ & $\mathfrak p$ & $\mathfrak k$ & $\mathfrak p$\\[2mm]
      & $H_{XXX}$ & $-$ & $\mathfrak p$ & $\mathfrak p$ & $\mathfrak k\cup\mathfrak p$ & $\mathfrak k\cup\mathfrak p$\\
      & $H_{XXY}$ & $-$ & $\mathfrak p$ & $\mathfrak p$ & $\mathfrak k\cup\mathfrak p$ & $\mathfrak k\cup\mathfrak p$\\
      & $H_{XYZ}$ & $-$ & $\mathfrak p$ & $\mathfrak p$ & $\mathfrak k\cup\mathfrak p$ & $\mathfrak k\cup\mathfrak p$\\[2mm]
\hline\\[-1mm]
$m=3$ & $H_{zzz}$ & $+$        & $\mathfrak k$ & $\mathfrak p$ & $\mathfrak k$ & $\mathfrak k$\\
      & $H_{xxx\pm yyy}$ & $+$ & $\mathfrak k$ & $\mathfrak k\cup\mathfrak p$ & $\mathfrak k$ & $\mathfrak p$\\
      & $H_{xxx + yyy + zzz}$ & $+$ & $\mathfrak k$ & $\mathfrak k\cup\mathfrak p$ & $\mathfrak k$ & $\mathfrak k\cup\mathfrak p$\\[2mm]
\hline\hline\\[-1mm]
\multicolumn{7}{l}{Note:}\\
\multicolumn{7}{l}{$\theta_{\rm CC}$ and $\theta_{\rm A\,II}$ are equivalent up to {\em non-local} permutation;}\\
\multicolumn{7}{l}{the same holds for $\theta_{\rm A\,II}$ and $\theta_{\rm A\,III}$ in the case of $p=q$.}
\end{tabular}\hspace{15mm}
\end{center}
\end{table}

\subsection*{Gradient Flows for Type-I Inversion by Local Unitaries}

Finally there is---fortunately---a convenient numerical solution to
the decision problem whether a given Hamiltonian $H$ generates a locally
invertible unitary, which is particularly helpful in cases where algebraic assessment is tedious.
It recasts the problem to the question whether the minimum of the distance
\begin{equation}
\Delta(K) := \norm{K H K^{-1} + H}_2^{\phantom 2} \equiv \norm{\Ad K (H) + H}_2^{\phantom 2}
\end{equation}
over all local unitaries $K\in\mathbf{SU}(2)^{\otimes n}$ is zero or not:
clearly, the norm ensures that $\Delta(K) = 0$ if and only if 
$\Ad K (H) = - H$,
which means \footnote{by $\norm{K H K^{-1} + H}_2^2 = 2\,\norm{H}^2 - 2\,\Re \tr\{K H K^{-1} (-H) \}$,
	where for hermitian $H$, the trace contains nothing but the real part}
\begin{equation}
f(K):= \Re \tr\{K H K^{-1} (-H) \}
\end{equation}
shall attain a global maximum that has to coincide with the upper bound for hermitian $H$  
reaching equality in $f_{\max}(K) \leq ||H||_2^2$.
Whether this limit can be reached by local unitaries may readlily be checked
numerically. To this end, one may devise a gradient flow along the lines of
Ref.~\cite{Science98,NMRJOGO} where, however, the gradient in the tangent space
has to be restricted by projecting it onto the algebra of local unitaries $\mathfrak k$
generating $\mathbf{SU}(2)^{\otimes n}$. As will be described elsewhere, establishing convergence
and appropriate step sizes of the iterative numerical scheme can be handled on a very
general level.
\begin{figure}[Ht!]
\includegraphics[scale=0.45]{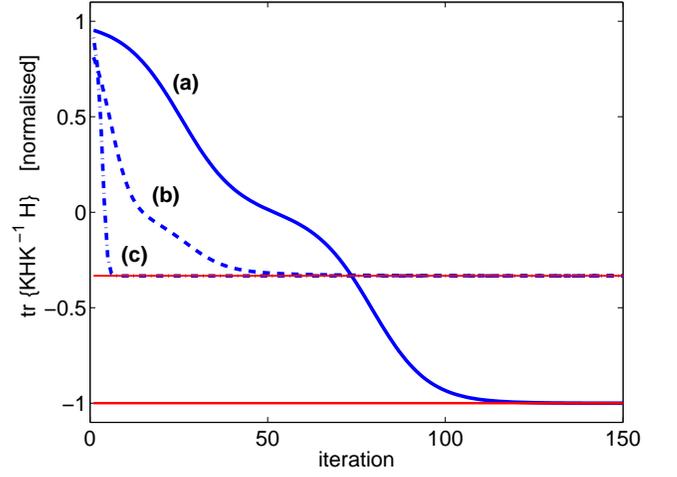}
\caption{\label{fig:ZZ_topo} (Colour online)
Gradient-flow driven local inversion of different Heisenberg interaction Hamiltonians:
(a) the $ZZ$ interaction on a cyclic four-qubit topology $C_4$ can be locally inverted,
(b) the $ZZ$ interaction on a cyclic three-qubit topology $C_3$ cannot be inverted locally,
(c) nor the $XXX$ isotropic interaction between two qubits.
}
\end{figure}
\begin{figure}[Ht!]
\includegraphics[scale=0.45]{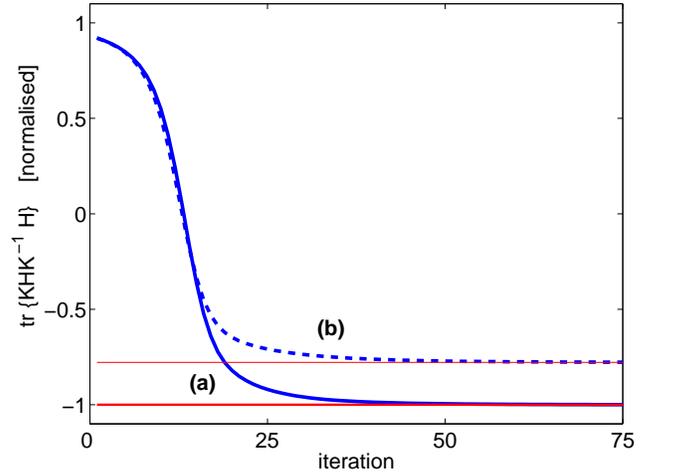}
\caption{\label{fig:C3_XX} (Colour online)
Gradient-flow driven local inversion of Heisenberg interactions of different quantum order
on a cyclic three-qubit topology $C_3$:
(a) the double-quantum interaction $X(-X)$ can be locally inverted, while
(b) the zero-quantum analogue $XX$ cannot.
}
\end{figure}

In the present context notice the gradient flow on the local unitaries takes the form
\begin{equation}
\begin{split}
\dot{K}\, = \,\grad f(K) 
	&= P_{\mathfrak{k}} ([(K H K^{-1}), H ])\; K\\[2mm]
	&= -P_{\mathfrak{k}} \big(\ad H\; \circ\;\Ad K (H)\,\big)\; K\;,
\end{split}
\end{equation}
where $P_{\mathfrak{k}}$ denotes the projection onto the
subalgebra $\mathfrak k$ of generators of local unitaries ${\mathbf K}=\mathbf{SU}(2)^{\otimes n}$.
The flow clearly reaches a critical point if already the entire commutator vanishes
\begin{equation}
[\Ad K (H), H] = 0
\quad .
\end{equation}
For hermitian $H$, this is the case, for instance whenever
\begin{equation}
\Ad K (H) = e^{\pm i p \pi} \;H 
\quad \text{(with $p=0,1,2,\dots$)}
\quad,
\end{equation} 
which means $H$ is an eigenoperator to $\Ad K (H)$, i.e.,
${\rm vec}\,H$ is an eigenvector of $(K^*\,\otimes K)$.
Eigenvectors $H$ to the eigenvalue $+1$ lead to global minima of $f(K)$, 
while global maxima are reached by eigenvectors $H$ to the eigenvalue $-1$.

In Figs.~\ref{fig:ZZ_topo} and \ref{fig:C3_XX}, we give some examples. Let $H$ be normalised to ${||H||}_2=1$.
If
$\tr \{KHK^{-1}H\} = -1$ can be reached, the interaction Hamiltonian
is locally invertible as in the case of the Heisenberg $ZZ$ interaction
in a cyclic four-qubit coupling topology (which clearly is a bipartite graph), 
while in the cyclic three-qubit
topology (obviously not forming a bipartite coupling graph) 
or in the case of the isotropic $XXX$ interaction it is not.

\subsection*{Relation to Local $C$-Numerical Ranges}
The $C$-numerical range is well-known \cite{Li94} to consist of the following set of
points in the complex plane
\begin{equation}
 W(C,A):=
        \{\tr\,(C^\dagger UAU^{-1}) \;|\; U \in SU(2^n)\} \;.
\end{equation}
In Ref.~\cite{loc_WCA}, we defined as 
{\em local $C$-numerical range} its subset
\begin{equation}
 W_{\rm{loc}}(C,A):=
        \{\tr\,(C^\dagger KAK^{-1}) \;|\; K \in SU(2)^{\otimes n}\} \;.
\end{equation}
In view of locally reversible Hamiltonians, things specialise to
$C=-H=-A$. Normalising again to $\norm H =1$, a locally reversible
Hamiltonian $H$ clearly requires $-1 \in W_{\rm{loc}}(-H,H)$.
This has just been exemplified by the numerical examples in the previous
section. Moreover, being a linear map of the local unitary orbit, 
the local $C$-numerical range is connected. 
For locally reversible $H$, one finds the real line segment 
$[-1;+1]=W_{\rm{loc}}(-H,H)$, whereas in Hamiltonians that fail to be
locally reversible, the line segment falls short of extending from $+1$ 
(which trivially always can be attained) to $-1$.

With these observations, the different aspects may be summed up.

\subsection*{Synopsis on Type-I Inversion}

\vspace{1mm}
\begin{corollary}[Local Time Reversal]
For an interaction Hamiltonian $H=H^\dagger$ with $\fnorm H = 1$ the following are equivalent:
\begin{enumerate}
\vspace{2mm}
\item $H$ is locally sign-reversible of type-I;
\vspace{2mm}
\item its {local $C$-numerical range} comprises $-1$:
        \qquad$ -1 \in W_{\rm loc}(-H,H)$\,;
\vspace{2mm}
\item its {local $C$-numerical range} is the real line segment from $-1$ to $+1$:
        \; $W_{\rm loc}(-H,H)\;=\;[-1\,;\,+1]$;
\vspace{2mm}
\item $\exists K\in SU(2)^{\otimes n}:$\\\qquad$\fnormsq{KHK^{-1} + H} = 0 \;\Leftrightarrow\; \Adr_K(H) = -H$
\vspace{2mm}
\item $H$ is locally unitarily similar to a $\Bar{H\;}$ with\\ $\Adr_{K_z}(\Bar{H\;}) = -\Bar{H\;}$;
\vspace{2mm}
\item let $\mathfrak g = \mathfrak g_0 \oplus\;\bigoplus\limits_{i\neq j} \C{} E_{ij}$
        be the root-space decomposition of $\mathfrak{sl}(N,\C{})$;
        $H$ is locally unitarily similar to a 
	{linear combination of root-space elements to non-zero roots}
	\begin{equation*}
        \Bar{H\;} := \sum\limits_{\lambda=1}^m c_\lambda E_{ij}^{(\lambda)}
	\end{equation*}
        satisfying a system of linear equations
	\begin{equation*}
        \sum_\ell p_{\lambda,\ell}\cdot\phi_\ell = \pm \pi(\mod 2 \pi)
	\end{equation*}
	in the sense of Eqn.~\ref{eqn:linsys}, where the $p_{\lambda,\ell}$
	can be interpreted as the quantum orders of the constituting 
	spherical tensor elements.

\end{enumerate}
\end{corollary}

\noindent
{\bf Proof:} The equivalence of (1) with statements (2) through (6) was of course already
proven in the respective sections.\\ Moreover, one finds
(1) $\Rightarrow$ (2): obvious;
(2) $\Rightarrow$ (3): connectedness of $W_{\rm loc}(C,A)$;
(3) $\Rightarrow$ (4): obvious;
(4) $\Rightarrow$ (5): Corollary~1;
(5) $\Rightarrow$ (6): Corollary~1, Proposition~1 and 2; as well as Corollary~2 for the interpretation
			as quantum orders;
(6) $\Rightarrow$ (1): Proposition~2.  \hfill$\blacksquare$

\begin{figure}[Ht!]
\includegraphics[width=0.95\columnwidth]{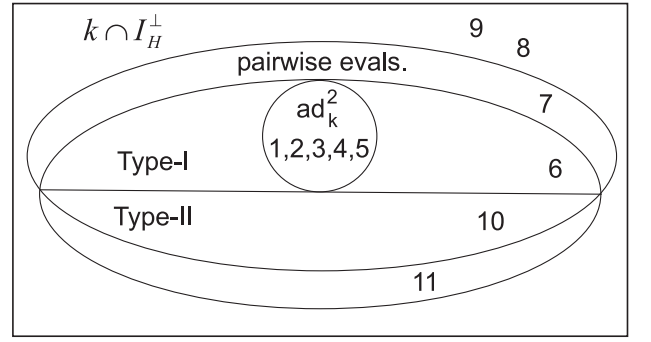}\caption{\label{fig:venn_inv} 
{\sc Venn} diagram showing that simple criteria of non-zero eigenvalues in pairs of opposite sign,
$\adr_k^2(H)=H$, and $I_H^{\perp}\cap\mathfrak{k}\neq\{\}$
fail to decide type-I invertibility as explained in the text.
The numbers in the sets refer to the examples listed in Tab.~\ref{tab:synopsis}.
}
\end{figure}
However, the simple necessary criteria of (i) non-zero eigenvalues occuring
in pairs of opposite sign, (ii) the intersection of the orthocomplement to the invariant subspace
with the generators of local unitaries not being empty $I_H^{\perp}\cap\mathfrak{k}\neq\{\}$, 
as well as the sufficient condition (iii) of the double commutator reproducing the Hamiltonian
in question, $\adr_k^2(H)=H$, fall short of giving a conclusive decision on type-I invertibility,
see Fig.~\ref{fig:venn_inv}.

\section{Pointwise Locally Invertible Propagators}
\begin{table}[Ht!]
\caption{\label{tab:synopsis}
Examples of pair interaction and multi-qubit interaction
Hamiltonians used to show the coverage by simple
type-I inversion criteria in Fig.~\ref{fig:venn_inv}.
}
\begin{ruledtabular}
\begin{tabular}{c|l}
Example& Hamiltonian \\
\hline 
1&$zz$\\        
2&$xx+yy$\\
3&$xx1-yy1 + x1x$\\
 &$-y1y + 1xx-1yy$\\
\\
4&$xx11-yy11-1xxx+1yxy+1yyx+1xyy$\\
 &$-x1xx+y1xy+y1yx+x1yy$\\
\\
5&$xx11-yy11 + x111-1xxx+1yxy+1yyx+1xyy$\\
 &$-x1xx+y1xy+y1yx+x1yy$\\
\\
6&$z11-xxx+xyy+yxy+yyx$\\
\\
7&$xx1+yy1+zz1 $\\
 &$- (1xx+1yy+1zz)$\\
\\
8&$xx1+yy1 + x1x$\\
 &$+y1y + 1xx+1yy$\\
\\
9&$zz1+z1z+1zz$\\
10&$zz+z1+1x$\\
11&$zz+z1+1z$\\
\end{tabular}
\end{ruledtabular}
\end{table}

Propagators that are not jointly invertible by a local unitary for all $t\in\R{}$ 
(together with the entire one-parameter group generated by their Hamiltonian)
may still be pointwise locally invertible at certain times $\tau$. 
So the task in this section is the following:
given some $\tau>0$, determine whether there is a pair 
$\{K_1,K_2\neq K_1^{-1}\} \subset \mathbf{K}:=\mathbf{SU}(2)^{\otimes n}\;$ 
so that
\begin{equation}\label{eqn:K1K2_super}
\begin{split}
K_1\; e^{-i\tau H}\; K_2 &= e^{+i\tau H}\\
\Leftrightarrow\; (K_2^t\otimes K_1^{\phantom{t}})\; \vec(e^{-i\tau H}) &= \vec(e^{+i\tau H})\;.
\end{split}
\end{equation}

\begin{remark}
Note that type-II invertibility only arises upon restriction 
to local operations $K_1, K_2 \in \mathbf K$, because
to any $U_0\in \mathbf{SU}(N)$ there is a trivial pair $U_1, U_2\in \mathbf{SU}(N)$
with $U_1 U_0 U_2 = U_0^{-1}$ (e.g. $U_1=U_0^{-2}, U_2=\unity$), 
whereas with $K_1, K_2 \in \mathbf K$
there is no such trivial generic solution unless $U_0\in \mathbf K$.
\end{remark}

\begin{corollary}
Let $H$ generate a one-parameter unitary group $\mathcal U:= \{e^{-it\,H} \,|\, t \in \R{}, iH \in \mathfrak{su}(N,\C{})\}$
that is locally invertible of type-I. Then
\begin{enumerate}
\item the generic elements of the left and right cosets 
	$\mathbf K \mathcal U$ and $\mathcal U \mathbf K$ 
	are type-II locally invertible;
\item in turn, every Hamiltonian that is type-II invertible is
	an element of a coset $\mathbf K \mathcal U$ or $\mathcal U \mathbf K$,
	where $\mathcal U$ is some one-parameter unitary group that itself is type-I invertible.
\end{enumerate}
\end{corollary}

Therefore type-II invertible propagators are a natural extension of the type"~I
invertible unitary one-parameter groups. In turn, however, 
the decision problem whether a given propagator is type"~II invertible
is generally quite complicated so that we will devise a coupled gradient flow on two local unitaries
for solving it numerically. Yet a number of cases can be treated algebraically
by analysing the symmetries of the matrix representation of the unitary
propagator to be inverted. 

Since these symmetry considerations
extend beyond the representation of unitary matrices,
we will ask whether an arbitrary given matrix can be mapped
to its hermitian adjoint by a superoperator of the form 
$(K_2^t\otimes K_1^{\phantom{t}})$
with local unitary $K_1, K_2$ (cp. Eqn.~\ref{eqn:K1K2_super}). 
To this end, one has to maximise the coincidences between 
$(K_2^t\otimes K_1^{\phantom{t}})$ and the adjoining superoperator
denoted $\widehat\Adj$ that takes its argument to the hermitian adjoint
(i.e. the complex conjugate transpose). Clearly, there is no local unitary
$(K_2^t\otimes K_1^{\phantom{t}})$ that fully matches with
$\widehat\Adj$ as this would be a universal inverting operator. 
However, there are classes of partial overlaps, where the lack of coincidence
enforces a symmetry in the matrices to be adjoined. These will be
analysed in detail in the following.

Because $\widehat\Adj$ 
has no matrix representation over the field of complex numbers, we
turn to the real domain. With $M_{\Re}$ and $M_{\Im}$ denoting the respective real and imaginary
parts of an arbitrary complex matrix $M$, one obtains a convenient representation
of $M$ as a real vector by virtue of the faithful mapping
\begin{equation}
M \mapsto \vec(M_{\Re}) \oplus \vec(M_{\Im})\;.
\end{equation}
[Note that this representation shows less redundance than the usual
$M \mapsto \left(\begin{smallmatrix} M_{\Re} \;-M_{\Im} \\ M_{\Im} %
	\;\phantom{-}M_{\Re} \end{smallmatrix}\right)$.]\\[-3mm]

In this notation, the adjoining superoperator does have a real matrix representation
defined via
\begin{equation}
\widehat{\Adj}_\R{} \big(\vec(M_{\Re})\oplus \vec(M_{\Im})\big) = \vec(M_{\Re}^t) \oplus \vec(-M_{\Im}^t)
\end{equation}
such as to take the form
\begin{equation}\label{eqn:Adj4_super}
\widehat\Adj_{\R{}} = \left(\begin{smallmatrix}
\widehat T & \phantom{-} \widehat 0 \\[1mm] \widehat 0 & - \widehat T \end{smallmatrix}\right)\quad,
\end{equation}
by virtue of the 
transposition superoperator $\widehat T$, which e.g. for the above representation of a 
$M\in\Mat_4(\C{})$ reads
\begin{equation}
\widehat T :=
\left(\,
\begin{smallmatrix}
\fvo & \cdt & \cdf & \cdf & \cdf \\
\cdf & \fvo & \cdt & \cdf & \cdf \\
\cdf & \cdf & \fvo & \cdt & \cdf \\
\cdf & \cdf & \cdf & \fvo & \cdt \\
\cdo & \fvo & \cdw & \cdf & \cdf & \cdf \\
\cdf & \cdo & \fvo & \cdw & \cdf & \cdf \\
\cdf & \cdf & \cdo & \fvo & \cdw & \cdf \\
\cdf & \cdf & \cdf & \cdo & \fvo & \cdw \\
\cdw & \fvo & \cdo & \cdf & \cdf & \cdf \\
\cdf & \cdw & \fvo & \cdo & \cdf & \cdf \\
\cdf & \cdf & \cdw & \fvo & \cdo & \cdf \\
\cdf & \cdf & \cdf & \cdw & \fvo & \cdo \\
\cdt & \fvo & \cdf & \cdf & \cdf \\
\cdf & \cdt & \fvo & \cdf & \cdf \\
\cdf & \cdf & \cdt & \fvo & \cdf \\
\cdf & \cdf & \cdf & \cdt & \fvo \\
\\
\end{smallmatrix}\,
\right)\quad.
\end{equation}

Likewise, for the local unitary transform
$K_1 M K_2 \;{\widehat=}\; \widehat K \vec(M)$
with $\widehat{K} := K_2^t\otimes K_1^{\phantom{t}}$
one gets the corresponding real representation of the superoperator via
\begin{equation}
\widehat{K}_{\R{}}\vec(M)_{\R{}} := \left(\begin{smallmatrix}
                \widehat K_{\Re} & -\widehat K_{\Im} \\[1mm] \widehat K_{\Im} & {\phantom{-}}\widehat K_{\Re}
                                                        \end{smallmatrix}\right)
                \big(\vec(M_{\Re})\oplus\vec(M_{\Im})\big)\;.
\end{equation}
Comparing the structure of 
$\widehat{K}_{\R{}}$ 
here and 
$\widehat\Adj_{\R{}}$ 
(in Eqn.~\ref{eqn:Adj4_super})
immediately shows
that for maximal coincidence the imaginary block $\widehat{K}_{\Im}$ within $\widehat{K}_{\R{}}$
has to vanish, because the row and column norms are limited to unity in (local) unitaries.
One may readily express the utmost possible overlaps of
$\widehat{K}_{\R{}}$ 
and
$\widehat\Adj_{\R{}}$ 
by taking the element\-wise Hadamard product as the coincidence matrix $\widehat C$
\begin{equation}
\begin{split}
\widehat C_{\R{}} :&= \widehat K_{\R{}} \odot \widehat\Adj_{\R{}} \\[3mm]
		&= \left(\begin{smallmatrix}
\widehat K_{\Re} & -\widehat K_{\Im} \\[1mm] \widehat K_{\Im} & {\phantom{-}}\widehat K_{\Re}
                                                        \end{smallmatrix}\right)
		\odot
		\left(\begin{smallmatrix} \widehat T & {\phantom{-}}\widehat 0 \\[1mm] \widehat 0 & -\widehat T \end{smallmatrix}\right) \\[3mm]
	&=
	\left(\begin{smallmatrix}
	\widehat C & \phantom{-} \widehat 0 \\[1mm] \widehat 0 & - \widehat C \end{smallmatrix}\right)\;,
\end{split}
\end{equation}
where $\widehat C$ reads, e.g. in the case $n=4$ 
\begin{equation}
\widehat C := \pm
\begin{small}
\left(
\begin{smallmatrix}
\fvn{a} & \cdt & \cdf & \cdf & \cdf \\
\cdf & \fvp{b} & \cdt & \cdf & \cdf \\
\cdf & \cdf & \fvp{c} & \cdt & \cdf \\
\cdf & \cdf & \cdf & \fvn{d} & \cdt \\
\cdo & \fvn{b} & \cdw & \cdf & \cdf & \cdf \\
\cdf & \cdo & \fvp{a} & \cdw & \cdf & \cdf \\
\cdf & \cdf & \cdo & \fvp{d} & \cdw & \cdf \\
\cdf & \cdf & \cdf & \cdo & \fvn{c} & \cdw \\
\cdw & \fvn{c} & \cdo & \cdf & \cdf & \cdf \\
\cdf & \cdw & \fvp{d} & \cdo & \cdf & \cdf \\
\cdf & \cdf & \cdw & \fvp{a} & \cdo & \cdf \\
\cdf & \cdf & \cdf & \cdw & \fvn{b} & \cdo \\
\cdt & \fvn{d} & \cdf & \cdf & \cdf \\
\cdf & \cdt & \fvp{c} & \cdf & \cdf \\
\cdf & \cdf & \cdt & \fvp{b} & \cdf \\
\cdf & \cdf & \cdf & \cdt & \fvn{a} \\
\\
\end{smallmatrix}
\right)
\end{small}
\end{equation}
and in which either $a$ or $b$ or $c$ or $d$ is unity.
Thus in the case $n=4$ one finds four subtypes of maximal overlap, termed $A, B, C, D$ henceforth.
According to the possible choices of signs, each of them occurs in four sign patterns
expressed by the indices $\nu\in\{A_{++++}, A_{+--+}, A_{-++-}, A_{----}\}$ and analogously for
subtypes $B,C,D$.

For instance,
let $\nu = A_{+--+}$,
then the local unitary $\widehat K_{\R{}}$ for maximal overlap 
with $\widehat\Adj_{\R{}}$ shows the following non-zero block $\widehat K_{\Re}$:
\begin{equation}
\widehat K_{\Re}^{(A_{+--+})} = 
\left(
\begin{smallmatrix}
\\
\Rfn & \cdt & \cdf & \cdf & \cdf\\
\cdo & \fwi & \cdw & \cdf & \cdf & \cdf\\
\cdw & \fwn & \cdo & \cdf & \cdf & \cdf\\
\cdt & \fwi & \cdf & \cdf & \cdf\\
\cdf & \fwn & \cdt & \cdf & \cdf\\
\cdf & \cdo & \Rfi & \cdw & \cdf & \cdf\\
\cdf & \cdw & \fwn & \cdo & \cdf & \cdf\\
\cdf & \cdt & \fwi & \cdf & \cdf\\
\cdf & \cdf & \fwi & \cdt & \cdf\\
\cdf & \cdf & \cdo & \fwn & \cdw & \cdf\\
\cdf & \cdf & \cdw & \Rfi & \cdo & \cdf\\
\cdf & \cdf & \cdt & \fwn & \cdf\\
\cdf & \cdf & \cdf & \fwi & \cdt\\
\cdf & \cdf & \cdf & \cdo & \fwn & \cdw\\
\cdf & \cdf & \cdf & \cdw & \fwi & \cdo\\
\cdf & \cdf & \cdf & \cdt & \Rfn\\
\\
\end{smallmatrix}
\right)\;,
\end{equation}
where the elements $\pm$ stand for $\pm 1$, while $*$ are unavoidable non-zero elements enforced by 
$\widehat K_{\R{}}$ being a local unitary. They do not contribute to $\widehat\Adj_{\R{}}$, 
in contrary, they bring about
unwanted actions on the argument, i.e. the matrix $M\in\Mat_n(\C{})$. Together with the lacking
elements for full overlap with $\widehat T$, they require the following symmetry in the matrix
argument 
\begin{equation}
M_{A_{+--+}} =
\left(
\begin{smallmatrix}
\\
  M_{11}^{\Re} &  0            &  0            &  M_{14}\\
  0            &  M_{22}^{\Im} &  M_{23}       &  0\\
  0            &  \pm M_{23}^* &  M_{33}^{\Im} &  0\\
\mp M_{14}^*   &  0            &  0            &  M_{44}^{\Re}
\\
\end{smallmatrix}
\right)
\end{equation}
in order to fulfill $\widehat K_{\R{}} \vec(M) = \vec(M^\dagger)$ as desired.

For the sake of completeness in the case of $n=4$, we give the remainder 
of constituents in the subtypes $A,B,C,D$ as well
as the associated sign patterns in Appendix"~C and D.

The structures of the pertinent block matrices $\widehat T$, $\widehat \Adj_{\R{}}$ and hence
$\widehat C$ are easily scalable to larger $n$: in Tab.~\ref{tab:K1K2-scaling} we give the number
of subtypes of coincidence as well as the number of different symmetry subtypes and sign patterns
in the matrix arguments $M\in\Mat_n(\C{})$ with growing number of dimensions.
\begin{table}[Ht!]
\begin{center}
\caption{Subtypes and Sign Patterns in Type-II Inversion}\label{tab:K1K2-scaling}
\begin{tabular}{ccccc}
\hline\hline\\[-1mm]
$\#$ spin qubits &\phantom{uu}& $\#$ subtypes &\phantom{uuuu}& $\#$ sign patterns\\[2mm]
\hline\\[-1mm]
$2$ && $4$ && $4$ \\[2mm]
$3$ && $8$ && $8$ \\[2mm]
$\vdots$ && $\vdots$ && $\vdots$ \\[2mm]
$n$ && $2^n$ && $2^n$ \\[2mm]
\hline\hline
\end{tabular}\hspace{15mm}
\end{center}
\end{table}

\subsection*{Type-II Inversion via Coupled Gradient Flows on Two Local Unitaries}
\begin{figure}[Ht!]
\includegraphics[scale=0.4]{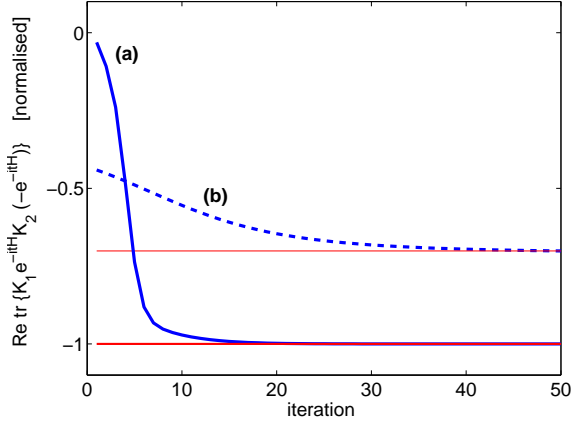}
\caption{\label{fig:invUV} (Colour online)
Gradient-flow driven local inversion of 
$U(t):=\exp\{-i\tfrac{\pi}{4}\;H\}$
with
$H = (\sigma_z\otimes\unity + \unity\otimes\sigma_z + \sigma_z\otimes\sigma_z)/2$
\\
(a) by coupled gradient flows on independent $K_1$ and $K_2$ and
(b) by a gradient flow with $K_2^{\protect\phantom{1}}=K_1^{-1}$.%
}
\end{figure}

In the general case, one may conveniently restate the problem of pointwise local invertibility
to the question,
whether for a fixed non-zero $\tau\in\R{}$ there is a pair $K_1,K_2\in {\mathbf K}=SU(2)^{\otimes n}$
so that
\begin{equation}
\begin{split}
K_1 e^{-i\tau H} K_2 &= e^{+i\tau H}\\
\Leftrightarrow
||K_1 e^{-i\tau H} K_2 - e^{+i\tau H}||_2 &= 0\quad.
\end{split}
\end{equation}
Then one may devise a coupled gradient flow on two local unitaries 
simultaneously in order to minimise
\begin{equation}
f(K_1,K_2) = \Re \;\tr\{K_1e^{-i\tau H}K_2(-e^{-i\tau H})\}
\end{equation}
by
(writing $U:=e^{-i\tau H}$ for short)
\begin{eqnarray}
\dot K_1 &=& \grad f(K_1) = P_{\mathfrak k}\,\big(K_1UK_2(-U)\big)\;K_1 \\
\dot K_2 &=& \grad f(K_2) = P_{\mathfrak k}\,\big(K_2(-U)K_1U\big)\;K_2 \quad.
\end{eqnarray}
Again, if $\tfrac{1}{N}\;\Re\;\tr\{K_1e^{-i\tau H}K_2(-e^{-i\tau H})\} = -1$
can be reached, then
$U(\tau)=e^{-i\tau H}$ is locally invertible at the point $\tau$. Examples are shown in 
Fig.~\ref{fig:invUV}.

\subsection*{Conclusion}

Generalising in the sense of Hahn's spin echo, we have characterised
all those effective multi-qubit quantum interactions allowing for time reversal
by manipulations confined to local unitary operations.
The evolutions generated by these interaction Hamiltonians
can be reversed and refocussed solely
by local unitaries. To this end, we have given a number
of necessary and sufficient conditions in terms of geometry,
eigenoperators, graphs of coupling topology, tensor analysis and
root-space decomposition.
Moreover, we have classified locally invertible evolutions into
two types. {\bf Type-I} consists of one-parameter groups generated by
Hamiltonians $H$ that are eigenoperators of local unitary conjugation
associated with the eigenvalue $-1$, i.e. $\Adr_K(H) = -H$.
We have shown how to construct the corresponding eigenspace
in closed algebraic form. Hamiltonians generating type"~I invertible
evolutions are locally unitarily similar to those invertible solely by
local $z$"~rotations, which thus can be regarded as normal form.
Taking the differential, we showed their components $E_{ij}$ relate to the
non-zero roots of the root-space decomposition of $\mathfrak{sl}(N,\C{})$
via $\adr_{k_z}(E_{ij})= (e_i-e_j)(E_{ij})$, where $k_z$ are the
generators of local $z$"~rotations. Moreover, in the special case
of joint local $z$"~rotations generated by $F_z$, the non-zero
roots were further shown to relate to the spherical tensors $T_{j,p}$ of 
non-zero quantum order $p$.
For Ising ZZ-coupling interactions as well as for Heisenberg $XY$ interactions
to be locally invertible of type-I, their coupling topology has to take the form of a bipartite graph.
An exception is the Heisenberg $X(-X)$ interaction, which is type-I invertible 
on any coupling topology, while Heisenberg $XXX$, $XXY$, and $XYZ$
interactions are not locally invertible at all because they relate
to rank-$0$ tensors, and their non-zero eigenvalues do not occur 
in pairs of opposite sign.

The pointwise invertible quantum evolutions of {\bf type-II}
are generalisations of those of type-I. They
consist of coset elements $\mathbf K \mathcal{U}$ and $\mathcal{U} \mathbf K$,
where $\mathcal{U}$ are type-I invertible one-parameter groups.
Here we have caracterised type-II invertible propagators by the symmetries
of their matrix representations.

Finally, in view of convenience in practical applications, we have devised
gradient-flow based numerical checks to decide
whether 
$$\underset{K_1, K_2\in SU(2)^{\otimes n}}{\min} \norm{K_1 e^{-itH} K_2 - e^{+itH}}^2_2 = 0\;,$$
i.e. whether a propagator is locally invertible of type-I or type-II or not.

\section{Appendix}
\subsection{Classification}
First, we prove {\bf Lemma~0} from the introduction:\\
Either $e^{-itH}$ is 
\begin{enumerate}
\item not invertible by local unitaries at all, or
\item it is trivial and self-inverse, or 
\item it is type-I invertible in the sense 
$\exists K\in SU(2)^{\otimes n}: KHK^{-1}=-H$ so $K e^{-itH} K^{-1} = e^{+itH}$ jointly for all $t\in\R{}$, 
or 
\item it is type-II invertible such that at some (but not all) points $\tau$ in time 
$K_1 e^{-i\tau H} K_2 = e^{+i\tau H}$ with $K_1, K_2 \in SU(2)^{\otimes n}$
and $K_2^{\phantom{1}}\neq K_1^{-1}$.
\end{enumerate}
{\bf Proof:} 
By the series expansion of the exponential one finds the obvious equivalence
\begin{equation}\label{eqn:affirm}
\begin{split}
KHK^{-1} &= -H\\
	&\Leftrightarrow \forall t \in \R{}: K e^{-itH} K^{-1} = e^{+itH}.
\end{split}
\end{equation}
Its logical negation 
\begin{equation}\label{eqn:negation}
\begin{split}
KHK^{-1} &\neq -H\\
	&\Leftrightarrow \exists t \in \R{}: K e^{-itH} K^{-1} \neq e^{+itH},
\end{split}
\end{equation}
comprises the following trivial cases
\begin{enumerate}
\item $\forall t \in \R{}: K e^{-itH} K^{-1} \neq e^{+itH}$, so either $e^{-itH}$
	is not locally invertible at all, or
\item $\exists \tau\neq 0: K e^{-i\tau H} K^{-1} = e^{+i\tau H}$, while for all other $t\neq\tau$
	(with exceptions of measure zero due to periodicity) $K e^{-itH} K^{-1} \neq e^{+itH}$ while
	$(KHK^{-1}) \neq -H$. This can only hold, if $K=\unity$  and
	$e^{-i\tau H}$ is self-inverse. 
\end{enumerate}
Otherwise, if the affirmative (Eqn~\ref{eqn:affirm}) is true one has
\begin{enumerate}
\item[3.] $\forall t \in \R{}: K e^{-itH} K^{-1} = e^{+itH}$.
\end{enumerate}
Finally, we have to show that type-I and II are distinct
\begin{enumerate}
\item[4.] $K_1 e^{-i\tau H} K_2 = e^{+i\tau H}$ with $K_1^{\phantom{1}}\neq K_2^{-1}$ may hold
	pointwise for certain $\tau$, but not for all $\tau\in\R{}$. Assume the contrary:
	$\forall t\in\R{}$ $K_1 e^{-it H} K_2 = e^{+it H}$ with $K_1^{\phantom{1}}\neq K_2^{-1}$
	and define as commuting elements of a one-parameter group
	$U_1:=e^{-it_1H}$ and $ U_2:=e^{-it_2H}$ to give $ U_{12}:=e^{-i(t_1+t_2)H}$. Then one has
	\begin{eqnarray*}
	K_1 U_{12} K_2 &=& U_{12}^{-1}\\
	K_1 U_1 U_2 K_2 &=& U_2^{-1} U_1^{-1} = U_1^{-1} U_2^{-1} \\
	(K_1 U_1 K_2)(\underline{K_2^{-1}} U_2^{\phantom{1}} K_2^{\phantom{1}}) &=& 
						(K_1 U_1 K_2)(\underline{K_1} U_2 K_2)\\
	K_2^{-1} &=& K_1^{\phantom{1}}\quad,
	\end{eqnarray*}
	where the latter contradicts the assumption. 
\end{enumerate}
These four instances prove Tab.~\ref{tab:classes}.
$\hfill\blacksquare$

\subsection{Explicit General Representation of $F_z$}
Recall that the generator of a joint $z$"~rotation on all the $n$ spin-$\tfrac{1}{2}$ qubits 
is defined as the diagonal matrix 
\begin{equation}
F_z:= \tfrac{1}{2}\sum\limits_{\ell=1}^n \unity^{(1)}\otimes\unity^{(2)}\otimes\cdots\otimes%
	\unity^{(\ell-1)}\otimes\sigma_z^{(\ell)}\otimes\unity^{(\ell+1)}\otimes\cdots\otimes\unity^{(n)}
\end{equation}
summing over the Pauli matrix $\sigma_z^{(\ell)}$ on all qubits.
Whenever it is necessary to express the total number of qubits, we write $F_z^n$.
Here we prove an explicit formula giving its $i^{\rm th}$ diagonal element for
general $n$.

\begin{lemma}
For ${(F_z^n)}_{ii}$ with the index $i\in\{1,2,3, \dots ,2^n\}$ calculate the
$n$-digit binary representation for the reduction by 1 as
$(i-1)_2=:\sum_{k=0}^{n-1}2^k b_k$. Then the $i^{\rm th}$ diagonal element
reads
\begin{equation}
{(F_z^n)}_{ii}=\tfrac{1}{2}\sum_{k=0}^{n-1}(-1)^{b_k}\quad.
\end{equation}

\end{lemma}

\noindent
{\bf Proof} (induction):\\
For $n=1$ one has:
$(i-1)_2=2^0b_0\in\{0,1\}$ so ${(F_z^1)}_{ii}=\frac{1}{2}\sum_{k=0}^{(1-1)}(-1)^{b_k}$
giving ${(F_z^1)}_{11}=\tfrac{1}{2}=-{(F_z^1)}_{22}$.\\

\noindent
In order to proceed from $n\rightarrow n+1$ we show that with ${(F_z^n)}_{ii}$ being given one finds
for the new index $i':=2^nb_n +i\in\{1,2,3, \dots, 2^{n+1}\}$
\begin{equation}
{(F_z^{n+1})}_{i'i'}=(\unity\otimes F_z^n)_{i'i'}+\tfrac{1}{2}(-1)^{b_n}\quad.
\end{equation}

\noindent
Use
\begin{equation}
\begin{split}
F_z^{n+1} &= \unity_2\otimes F_z^{n}+I_z\otimes\unity_{2^n}\\[2mm]
	  &= \diag\left({(F_z^n)}_{11},\dots,{(F_z^n)}_{2^n 2^n}; {(F_z^n)}_{11},\dots,{(F_z^n)}_{2^n 2^n}\right)\\
	  &\phantom{XX} +\tfrac{1}{2}\diag(1,1, \dots, 1;-1,-1, \dots,-1)
\end{split}
\end{equation}
to see that
the last term adds $\tfrac{1}{2}$ for $i'\in \{1,\dots,2^n\} =i$ 
and $-\tfrac{1}{2}$ for $i'\in \{2^n+1,\dots,2^{n+1}\} =2^n + i$,
in coincidence with $b_n$ taking the value $0$ or $1$. $\hfill\blacksquare$

\clearpage
\newpage
\subsection{Subtypes of Pointwise Invertible Local Unitaries}
For the case of two qubits, we give local unitary superoperators
of different type of partial overlap with the adjoining superoperator
\begin{equation}
\widehat{K}_{\R{}}\vec(M)_{\R{}} := \left(\begin{smallmatrix}
                \widehat K_{\Re} & -\widehat K_{\Im} \\[1mm] \widehat K_{\Im} & {\phantom{-}}\widehat K_{\Re}
                                                        \end{smallmatrix}\right)
                \big(\vec(M_{\Re})\oplus\vec(M_{\Im})\big)\;.
\end{equation}
In subtype~$A$, the blockmatrix $\widehat K_{\Re}$ within the above supermatrix $\widehat K_\R{}$ 
may take four different forms according to the indices
$$
A_{++++} =  -A_{----} = 
\left(
\begin{smallmatrix}
\\
\Rfn & \cdt & \cdf & \cdf & \cdf\\
\cdo & \fwi & \cdw & \cdf & \cdf & \cdf\\
\cdw & \fwn & \cdo & \cdf & \cdf & \cdf\\
\cdt & \fwi & \cdf & \cdf & \cdf\\
\cdf & \fwi & \cdt & \cdf & \cdf\\
\cdf & \cdo & \Rfn & \cdw & \cdf & \cdf\\
\cdf & \cdw & \fwi & \cdo & \cdf & \cdf\\
\cdf & \cdt & \fwn & \cdf & \cdf\\
\cdf & \cdf & \fwn & \cdt & \cdf\\
\cdf & \cdf & \cdo & \fwi & \cdw & \cdf\\
\cdf & \cdf & \cdw & \Rfn & \cdo & \cdf\\
\cdf & \cdf & \cdt & \fwi & \cdf\\
\cdf & \cdf & \cdf & \fwi & \cdt\\
\cdf & \cdf & \cdf & \cdo & \fwn & \cdw\\
\cdf & \cdf & \cdf & \cdw & \fwi & \cdo\\
\cdf & \cdf & \cdf & \cdt & \Rfn\\
\\
\end{smallmatrix}
\right)
$$
$$
A_{+--+} = -A_{-++-} = 
\left(
\begin{smallmatrix}
\\
\Rfn & \cdt & \cdf & \cdf & \cdf\\
\cdo & \fwi & \cdw & \cdf & \cdf & \cdf\\
\cdw & \fwn & \cdo & \cdf & \cdf & \cdf\\
\cdt & \fwi & \cdf & \cdf & \cdf\\
\cdf & \fwn & \cdt & \cdf & \cdf\\
\cdf & \cdo & \Rfi & \cdw & \cdf & \cdf\\
\cdf & \cdw & \fwn & \cdo & \cdf & \cdf\\
\cdf & \cdt & \fwi & \cdf & \cdf\\
\cdf & \cdf & \fwi & \cdt & \cdf\\
\cdf & \cdf & \cdo & \fwn & \cdw & \cdf\\
\cdf & \cdf & \cdw & \Rfi & \cdo & \cdf\\
\cdf & \cdf & \cdt & \fwn & \cdf\\
\cdf & \cdf & \cdf & \fwi & \cdt\\
\cdf & \cdf & \cdf & \cdo & \fwn & \cdw\\
\cdf & \cdf & \cdf & \cdw & \fwi & \cdo\\
\cdf & \cdf & \cdf & \cdt & \Rfn\\
\\
\end{smallmatrix}
\right)
$$
\noindent
Subtype~$B$ comprises the forms
$$
B_{++++} =  -B_{----} = 
\left(
\begin{smallmatrix}
\\
\cdw & \cdt & \fwi & \cdf & \cdw & \cdf\\
\cdf & \Rfn & \cdo & \cdw & \cdf & \cdf\\
\cdf & \cdt & \fwn & \cdf & \cdf\\
\cdf & \cdw & \fwi & \cdo & \cdf & \cdf\\
\cdo & \Rfn & \cdf & \cdf & \cdw & \cdf\\
\fwi & \cdo & \cdw & \cdf & \cdf & \cdf\\
\cdt & \fwi & \cdf & \cdf & \cdf\\
\cdw & \fwn & \cdo & \cdf & \cdf & \cdf\\
\cdf & \cdf & \cdw & \cdt & \fwn & \cdw\\
\cdf & \cdf & \cdf & \fwi & \cdo & \cdw\\
\cdf & \cdf & \cdf & \cdt & \fwi\\
\cdf & \cdf & \cdf & \cdw & \Rfn & \cdo\\
\cdf & \cdf & \cdo & \fwi & \cdf & \cdw\\
\cdf & \cdf & \fwn & \cdo & \cdw & \cdf\\
\cdf & \cdf & \cdt & \Rfn & \cdf\\
\cdf & \cdf & \cdw & \fwi & \cdo & \cdf\\
\\
\end{smallmatrix}
\right)
$$
$$
B_{+--+} = -B_{-++-} = 
\left(
\begin{smallmatrix}
\\
\cdw & \cdt & \fwn & \cdf & \cdw & \cdf\\
\cdf & \Rfi & \cdo & \cdw & \cdf & \cdf\\
\cdf & \cdt & \fwi & \cdf & \cdf\\
\cdf & \cdw & \fwn & \cdo & \cdf & \cdf\\
\cdo & \Rfn & \cdf & \cdf & \cdw & \cdf\\
\fwi & \cdo & \cdw & \cdf & \cdf & \cdf\\
\cdt & \fwi & \cdf & \cdf & \cdf\\
\cdw & \fwn & \cdo & \cdf & \cdf & \cdf\\
\cdf & \cdf & \cdw & \cdt & \fwn & \cdw\\
\cdf & \cdf & \cdf & \fwi & \cdo & \cdw\\
\cdf & \cdf & \cdf & \cdt & \fwi\\
\cdf & \cdf & \cdf & \cdw & \Rfn & \cdo\\
\cdf & \cdf & \cdo & \fwn & \cdf & \cdw\\
\cdf & \cdf & \fwi & \cdo & \cdw & \cdf\\
\cdf & \cdf & \cdt & \Rfi & \cdf\\
\cdf & \cdf & \cdw & \fwn & \cdo & \cdf\\
\\
\end{smallmatrix}
\right)
$$
\newpage
\noindent
Subtype~$C$ has the forms
$$
C_{++++} = -C_{----} = 
\left(
\begin{smallmatrix}
\\
\cdf & \cdf & \cdw & \fwi & \cdo & \cdf\\
\cdf & \cdf & \cdt & \fwn & \cdf\\
\cdf & \cdf & \Rfn & \cdo & \cdw & \cdf\\
\cdf & \cdw & \cdt & \fwi & \cdf & \cdw\\
\cdf & \cdf & \cdf & \cdw & \fwn & \cdo\\
\cdf & \cdf & \cdf & \cdt & \fwi\\
\cdf & \cdf & \cdf & \fwi & \cdo & \cdw\\
\cdf & \cdf & \cdw & \cdt & \Rfn & \cdw\\
\cdw & \Rfn & \cdo & \cdf & \cdf & \cdf\\
\cdt & \fwi & \cdf & \cdf & \cdf\\
\fwi & \cdo & \cdw & \cdf & \cdf & \cdf\\
\cdo & \fwn & \cdf & \cdf & \cdw & \cdf\\
\cdf & \cdw & \fwi & \cdo & \cdf & \cdf\\
\cdf & \cdt & \Rfn & \cdf & \cdf\\
\cdf & \fwn & \cdo & \cdw & \cdf & \cdf\\
\cdw & \cdt & \fwi & \cdf & \cdw & \cdf\\
\\
\end{smallmatrix}
\right)
$$
$$
C_{+--+} = -C_{-++-} = 
\left(
\begin{smallmatrix}
\\
\cdf & \cdf & \cdw & \fwi & \cdo & \cdf\\
\cdf & \cdf & \cdt & \fwi & \cdf\\
\cdf & \cdf & \Rfi & \cdo & \cdw & \cdf\\
\cdf & \cdw & \cdt & \fwi & \cdf & \cdw\\
\cdf & \cdf & \cdf & \cdw & \fwn & \cdo\\
\cdf & \cdf & \cdf & \cdt & \fwn\\
\cdf & \cdf & \cdf & \fwn & \cdo & \cdw\\
\cdf & \cdf & \cdw & \cdt & \Rfn & \cdw\\
\cdw & \Rfn & \cdo & \cdf & \cdf & \cdf\\
\cdt & \fwn & \cdf & \cdf & \cdf\\
\fwn & \cdo & \cdw & \cdf & \cdf & \cdf\\
\cdo & \fwn & \cdf & \cdf & \cdw & \cdf\\
\cdf & \cdw & \fwi & \cdo & \cdf & \cdf\\
\cdf & \cdt & \Rfi & \cdf & \cdf\\
\cdf & \fwi & \cdo & \cdw & \cdf & \cdf\\
\cdw & \cdt & \fwi & \cdf & \cdw & \cdf\\
\\
\end{smallmatrix}
\right)
$$
\noindent
Subtype~$D$ includes the forms
$$
D_{++++} = -D_{----} = 
\left(
\begin{smallmatrix}
\\
\cdf & \cdf & \cdf & \cdt & \fwi\\
\cdf & \cdf & \cdf & \cdw & \fwn & \cdo\\
\cdf & \cdf & \cdf & \cdo & \fwi & \cdw\\
\cdf & \cdf & \cdf & \Rfn & \cdt\\
\cdf & \cdf & \cdt & \fwn & \cdf\\
\cdf & \cdf & \cdw & \fwi & \cdo & \cdf\\
\cdf & \cdf & \cdo & \Rfn & \cdw & \cdf\\
\cdf & \cdf & \fwi & \cdt & \cdf\\
\cdf & \cdt & \fwi & \cdf & \cdf\\
\cdf & \cdw & \Rfn & \cdo & \cdf & \cdf\\
\cdf & \cdo & \fwi & \cdw & \cdf & \cdf\\
\cdf & \fwn & \cdt & \cdf & \cdf\\
\cdt & \Rfn & \cdf & \cdf & \cdf\\
\cdw & \fwi & \cdo & \cdf & \cdf & \cdf\\
\cdo & \fwn & \cdw & \cdf & \cdf & \cdf\\
\fwi & \cdt & \cdf & \cdf & \cdf\\
\\
\end{smallmatrix}
\right)
$$
$$
D_{+--+} = -D_{-++-} = 
\left(
\begin{smallmatrix}
\\
\cdf & \cdf & \cdf & \cdt & \fwn\\
\cdf & \cdf & \cdf & \cdw & \fwi & \cdo\\
\cdf & \cdf & \cdf & \cdo & \fwi & \cdw\\
\cdf & \cdf & \cdf & \Rfn & \cdt\\
\cdf & \cdf & \cdt & \fwn & \cdf\\
\cdf & \cdf & \cdw & \fwi & \cdo & \cdf\\
\cdf & \cdf & \cdo & \Rfi & \cdw & \cdf\\
\cdf & \cdf & \fwn & \cdt & \cdf\\
\cdf & \cdt & \fwn & \cdf & \cdf\\
\cdf & \cdw & \Rfi & \cdo & \cdf & \cdf\\
\cdf & \cdo & \fwi & \cdw & \cdf & \cdf\\
\cdf & \fwn & \cdt & \cdf & \cdf\\
\cdt & \Rfn & \cdf & \cdf & \cdf\\
\cdw & \fwi & \cdo & \cdf & \cdf & \cdf\\
\cdo & \fwi & \cdw & \cdf & \cdf & \cdf\\
\fwn & \cdt & \cdf & \cdf & \cdf\\
\\
\end{smallmatrix}
\right)
$$
\clearpage
\newpage
\enlargethispage{9mm}
\subsection{Symmetries in the Argument}
According to the classes of partial overlap with
the adjoining superoperator, here we give the according
symmetries for the matrices $M\in \Mat_4(\C{})$ to be
mapped to their adjoints by the corresponding local unitaries.
\noindent
\begin{eqnarray*}
M_{A_{++++}} &=&
\left(
\begin{smallmatrix}
\\
 M_{11}^{\Re} &  M_{12}       &  M_{13}       & M_{14}\\
 0            &  M_{22}^{\Re} &  M_{23}       & M_{24}\\
 0            &  0            &  M_{33}^{\Re} & M_{34}\\
 0            &  0            &  0            & M_{44}^{\Re}
\\
\end{smallmatrix}
\right)\\[2mm]
&& + (-1^{a\cdot b})
\left(
\begin{smallmatrix}
\\
 0            &  0            &  0             & \quad0\quad  \\[1.15mm]
-1^a M_{12}^* &  0            &  0             & 0     \\
-1^b M_{13}^* &  M_{23}^*     &  0             & 0     \\
 M_{14}^*     & -1^b M_{24}^* &  -1^a M_{34}^* & 0         
\\
\end{smallmatrix}
\right)\\[4mm]
M_{A_{+--+}} &=&
\left(
\begin{smallmatrix}
\\
  M_{11}^{\Re} &  0            &  0            &  M_{14}\\
  0            &  M_{22}^{\Im} &  M_{23}       &  0\\
  0            &  \pm M_{23}^* &  M_{33}^{\Im} &  0\\
\mp M_{14}^*   &  0            &  0            &  M_{44}^{\Re}
\\
\end{smallmatrix}
\right)\\[4mm]
M_{A_{-++-}} &=&
\left(
\begin{smallmatrix}
\\
  M_{11}^{\Im} &  0            &  0            &  M_{14}\\
  0            &  M_{22}^{\Re} &  M_{23}       &  0\\
  0            &  \pm M_{23}^* &  M_{33}^{\Re} &  0\\
\mp M_{14}^*   &  0            &  0            &  M_{44}^{\Im}
\\
\end{smallmatrix}
\right)\\[4mm]
M_{A_{----}} &=&
\left(
\begin{smallmatrix}
\\
 M_{11}^{\Im} &  M_{12}       &  M_{13}       & M_{14}\\
 0            &  M_{22}^{\Im} &  M_{23}       & M_{24}\\
 0            &  0            &  M_{33}^{\Im} & M_{34}\\
 0            &  0            &  0            & M_{44}^{\Im}
\\
\end{smallmatrix}
\right)\\[2mm]
&& - (-1^{a\cdot b})
\left(
\begin{smallmatrix}
\\
 0            &  0            &  0             & \quad0\quad     \\[1.15mm]
-1^a M_{12}^* &  0            &  0             & 0     \\
-1^b M_{13}^* &  M_{23}^*     &  0             & 0     \\
 M_{14}^*     & -1^b M_{24}^* &  -1^a M_{34}^* & 0         
\\
\end{smallmatrix}
\right)
\end{eqnarray*}
\HRule
\begin{eqnarray*}
M_{B_{++++}} &=&
\left(
\begin{smallmatrix}
\\
 M_{11}       &  M_{12}^{\Re} &  M_{13}       &  M_{14}\\
 M_{21}^{\Re} &  0            &  M_{23}       &  M_{24}\\
 0            &  0            &  M_{33}       &  M_{34}^{\Re} \\
 0            &  0            &  M_{43}^{\Re} &  0       
\\
\end{smallmatrix}
\right)\\[2mm]
&& + (-1^{a\cdot b})
\left(
\begin{smallmatrix}
\\
 0            &  0             &  \quad0\quad   &  0         \\[1.15mm]
 0            &  -1^a M_{11}^* &  0             &  0         \\
 M_{24}^*     & -1^b M_{14}^* &  0             &  0         \\
-1^b M_{23}^* &  M_{13}^*     &  0             &  -1^a M_{33}^*
\\
\end{smallmatrix}
\right)\\[4mm]
M_{B_{+--+}} &=&
\left(
\begin{smallmatrix}
\\
     0        & M_{12}^{\Im}   &  M_{13}          &      0\\
 M_{21}^{\Re} &       0        &  0               &   M_{24}\\
 \pm M_{24}^* &       0        &  0               &   M_{34}^{\Re}\\
     0        & \mp M_{13}^*   &  M_{43}^{\Im}    &      0
\\
\end{smallmatrix}
\right)\\[4mm]
M_{B_{-++-}} &=&
\left(
\begin{smallmatrix}
\\
     0        & M_{12}^{\Re}   &  M_{13}          &      0\\
 M_{21}^{\Im} &       0        &  0               &   M_{24}\\
 \pm M_{24}^* &       0        &  0               &   M_{34}^{\Im}\\
     0        & \mp M_{13}^*   &  M_{43}^{\Re}    &      0
\\
\end{smallmatrix}
\right)\\[4mm]
M_{B_{----}} &=&
\left(
\begin{smallmatrix}
\\
 M_{11}       &  M_{12}^{\Im} &  M_{13}       &  M_{14}\\
 M_{21}^{\Im} &  0            &  M_{23}       &  M_{24}\\
 0            &  0            &  M_{33}       &  M_{34}^{\Im} \\
 0            &  0            &  M_{43}^{\Im} &  0       
\\
\end{smallmatrix}
\right)\\[2mm]
&& - (-1^{a\cdot b})
\left(
\begin{smallmatrix}
\\
 0            &  0            &  \quad0\quad   &  0         \\[1.15mm]
 0            & -1^a M_{11}^* &  0             &  0         \\
 M_{24}^*     & -1^b M_{14}^* &  0             &  0         \\
-1^b M_{23}^* &  M_{13}^*     &  0             &  -1^a M_{33}^*
\\
\end{smallmatrix}
\right)
\end{eqnarray*}

\begin{eqnarray*}
M_{C_{++++}} &=&
\left(
\begin{smallmatrix}
\\
 M_{11}       &  M_{12}       &  M_{13}^{\Re} &  M_{14}\\
 M_{21}       &  M_{22}       &  0            &  M_{24}^{\Re}\\
 M_{24}^{\Re} &  M_{32}       &  0            &  0       \\
 0            &  M_{13}^{\Re} &  0            &  0        
\\
\end{smallmatrix}
\right)\\[2mm]
&& + (-1^{a\cdot b})
\left(
\begin{smallmatrix}
\\
     0        & \quad0\quad    &       0        &      0     \\[1.15mm]
     0        &       0        & -1^a  M_{14}^* &      0     \\
     0        &       0        & -1^b  M_{11}^* &      M_{21}^*\\
-1^a M_{32}^* &       0        &       M_{12}^* & -1^b M_{22}^*
\\
\end{smallmatrix}
\right)\\[4mm]
M_{C_{+--+}} &=&
\left(
\begin{smallmatrix}
\\
 0            &  M_{12}       &  M_{13}^{\Im} &  0     \\
 M_{21}       &  0            &  0            &  M_{24}^{\Re}\\
 M_{24}^{\Re} &  0            &  0            & \pm M_{21}^*\\
 0            &  M_{13}^{\Im} &  \mp M_{12}^*     &  0       
\\
\end{smallmatrix}
\right)\\[4mm]
M_{C_{-++-}} &=&
\left(
\begin{smallmatrix}
\\
 0            &  M_{12}       &  M_{13}^{\Re} &  0     \\
 M_{21}       &  0            &  0            &  M_{24}^{\Im}\\
 M_{24}^{\Im} &  0            &  0            & \pm M_{21}^*\\
 0            &  M_{13}^{\Re} &  \mp M_{12}^*     &  0       
\\
\end{smallmatrix}
\right)\\[4mm]
M_{C_{----}} &=&
\left(
\begin{smallmatrix}
\\
 M_{11}       &  M_{12}       &  M_{13}^{\Im} &  M_{14}\\
 M_{21}       &  M_{22}       &  0            &  M_{24}^{\Im}\\
 M_{24}^{\Im} &  M_{32}       &  0            &  0       \\
 0            &  M_{13}^{\Im} &  0            &  0        
\\
\end{smallmatrix}
\right)\\[2mm]
&& - (-1^{a\cdot b})
\left(
\begin{smallmatrix}
\\
     0        &  \quad0\quad   &       0        &      0     \\[1.15mm]
     0        &       0        & -1^a  M_{14}^* &      0     \\
     0        &       0        & -1^b  M_{11}^* &      M_{21}^*\\
-1^a M_{32}^* &       0        &       M_{12}^* & -1^b M_{22}^*
\\
\end{smallmatrix}
\right)
\end{eqnarray*}
\HRule

\begin{eqnarray*}
M_{D_{+++}} &=&
\left(
\begin{smallmatrix}
\\
 M_{11}       &   M_{12}       &  M_{13}       &  M_{14}^{\Re}  \\
 M_{21}       &   M_{22}       &  M_{23}^{\Re} &  0             \\
 M_{31}       &   M_{32}^{\Re} &  0            &  0             \\
 M_{41}^{\Re} &   0            &  0            &  0                    
\\
\end{smallmatrix}
\right)\\[2mm]
&& + (-1^{a\cdot b})
\left(
\begin{smallmatrix}
\\
\quad0\quad   &   0            &  0            &  0             \\[1.15mm]
 0            &   0            &  0            & -1^a M_{13}^*\\
 0            &   0            &  M_{22}^*     & -1^b M_{21}^*\\
 0            &  -1^a M_{31}^* & -1^b M_{21}^* &  M_{11}^*
\\
\end{smallmatrix}
\right)\\[4mm]
M_{D_{+--+}} &=&
\left(
\begin{smallmatrix}
\\
 M_{11}       &      0            &  0            &  M_{14}^{\Re}  \\
 0            &      M_{22}       &  M_{23}^{\Im} &  0       \\
 0            &      M_{32}^{\Im} & \pm M_{22}^*  &  0       \\
 M_{41}^{\Re} &      0            &  0            &  \mp M_{11}^*
\\
\end{smallmatrix}
\right)\\[4mm]
M_{D_{-++-}} &=&
\left(
\begin{smallmatrix}
\\
 M_{11}       &      0            &  0            &  M_{14}^{\Im}  \\
 0            &      M_{22}       &  M_{23}^{\Re} &  0       \\
 0            &      M_{32}^{\Re} & \pm M_{22}^*  &  0       \\
 M_{41}^{\Im} &      0            &  0            &  \mp M_{11}^*
\\
\end{smallmatrix}
\right)\\[4mm]
M_{D_{----}} &=&
\left(
\begin{smallmatrix}
\\
 M_{11}       &   M_{12}       &  M_{13}       &  M_{14}^{\Im}  \\
 M_{21}       &   M_{22}       &  M_{23}^{\Im} &  0             \\
 M_{31}       &   M_{32}^{\Im} &  0            &  0             \\
 M_{41}^{\Im} &   0            &  0            &  0                    
\\
\end{smallmatrix}
\right)\\[2mm]
&& - (-1^{a\cdot b})
\left(
\begin{smallmatrix}
\\
\quad0\quad   &   0            &  0            &  0             \\[1.15mm]
 0            &   0            &  0            & -1^a M_{13}^*\\
 0            &   0            &  M_{22}^*     & -1^b M_{21}^*\\
 0            &  -1^a M_{31}^* & -1^b M_{21}^* &  M_{11}^*
\\
\end{smallmatrix}
\right)
\end{eqnarray*}
\clearpage

\begin{acknowledgments}
We are indebted to Prof.~Steffen Glaser for useful comments and support.
Valuable discussions with Shashank Virmani (Imperial College, London)
on pointwise inversion as well as with Gunther Dirr (W{\"u}rzburg University)
on Cartan-like decomposition are gratefully acknowledged.
This work was supported in part by {\em Deutsche Forschungsgemeinschaft}, DFG,
within the incentive {\em \/`Quanteninformations\-verarbeitung\/'}, QIV as well as
by the integrated EU project QAP.
\end{acknowledgments}



\end{document}